\documentclass[a4paper,11pt]{article}
\usepackage{jheppub}

\hypersetup{
    colorlinks,
    linkcolor={red!50!black},
    citecolor={blue!50!black},
    urlcolor={blue!80!black}
  }
      
\usepackage[T1]{fontenc}
\DeclareSymbolFont{usualmathcal}{OMS}{cmsy}{m}{n}
\DeclareSymbolFontAlphabet{\mathcal}{usualmathcal}

\DeclareMathOperator{\Tr}{Tr\,}
\DeclareMathOperator{\tr}{tr\,}

\numberwithin{equation}{section}

\usepackage{graphicx}
\usepackage{hyperref}

\usepackage{tensor}
\usepackage[most]{tcolorbox}
\usepackage{slashed}
\usepackage{bm}

\newcommand{\p}{\partial}

\def\be{\begin{equation}} 
\def\ee{\end{equation}}

\def\de{\partial}

\def\a{\alpha} 
\def\b{\beta} 
\def\g{\gamma} 
\def\G{\Gamma} 
\def\D{\Delta} 
\def\d{\delta} 
 
\def\eps{\varepsilon}

\def\l{\lambda} 
\def\L{\Lambda}

\def\r{\rho} 
\def\s{\sigma} 
 
\def\t{\tau}

\def\w{\omega} 
\def\W{\Omega} 
 
\def\th{\theta}

\title{Hydrodynamics, anomaly inflow and bosonic effective field theory}

\author[a]{Alexander G. Abanov,}
\author[b,1]{Andrea Cappelli\note{Corresponding author.}}

\affiliation[a]{ Department of Physics and Astronomy,
Stony Brook University, Stony Brook, NY 11794, USA}
\affiliation[b]{INFN, Sezione di Firenze,
  Via G. Sansone, 1, 50019 Sesto Fiorentino, Firenze, Italy}

\emailAdd{alexandre.abanov@stonybrook.edu}
\emailAdd{andrea.cappelli@fi.infn.it}

\abstract{
Euler hydrodynamics of perfect fluids can be viewed as an
effective bosonic field theory. In  cases when the underlying
microscopic system involves Dirac fermions, the quantum anomalies
should be properly described. In 1+1 dimensions the
action formulation of hydrodynamics at zero temperature is
reconsidered and shown to be equal to standard field-theory bosonization.
Furthermore, it can be derived from a topological gauge theory in one extra
dimension, which identifies the fluid variables through the
anomaly inflow relations.
Extending this framework to 3+1 dimensions yields an effective field
theory/hydrodynamics model, capable of elucidating the mixed
axial-vector and axial-gravitational anomalies of Dirac fermions. This
formulation provides a platform for bosonization in higher
dimensions. Moreover, the connection with 4+1 dimensional topological
theories suggests some generalizations of fluid dynamics
involving additional degrees of freedom.}

\begin{document}
\maketitle
\flushbottom

%-1-%%%%%%%%%%%%%%%%%%%%%%%%%
\section{Introduction}
%%%%%%%%%%%%%%%%%%%%%%%%%%

In this study we aim to discuss the effective dynamics of Dirac
fermions and their anomalies in 1+1 and 3+1 dimensions. In particular,
we investigate the connections between the methods of 
bosonic field theory and hydrodynamics.

On one hand, bosonization, the relation between bosonic and fermionic
field theories, is now attracting much interest beyond the
well-understood 1+1 dimensional case
\cite{Alekseev,Fradkin-boso,Cappelli:2016xwp,Fradkin:2016ksx,
  Kapustin-bosonI,Kapustin-bosonII,Andreucci:2019ltx,Cappelli-vertex}.
On the other hand, hydrodynamic theories incorporating anomalous
currents are actively developed
\cite{Volovik-Nature,Son:2009tf,Loganayagam2011anomaly,Banerjee:2012iz,
  Nair:2011,Haehl:2013hoa,Jensen-transgr,
  Dubovsky:2011sk,monteiro2015hydrodynamics,Nair:2020kjg,AbanovI}.
Both approaches can be viewed as constructions of effective theories
for interacting fermionic systems.

In order to understand the relations between these themes we take advantage
of the studies on topological phases of matter, which can be described
in terms of both fermionic and bosonic degrees of freedom
\cite{Zhang-rev,Ludwig-rev,Witten-lect,Wen2019}.  Physical systems in these
phases usually possess a large bulk gap and massless
boundary excitations. Within the low-energy effective field theory
approach, the bulk is described by a topological gauge theory and the
boundary by a relativistic field theory which is anomalous. Bulk and
boundary theories are related by anomaly inflow and anomaly matching,
enforcing gauge symmetry in the combined bulk-edge system
\cite{arouca2022quantum}.

Topological states are primarily (interacting) fermionic systems, yet
there exists a well-developed bosonic description for both bulk and
boundary excitations. A notable example is the quantum Hall effect,
where the bulk is 2+1 dimensional and the boundary 1+1 dimensional
\cite{Wenbook,fradkinbook}. The respective degrees of freedom are
interconnected: the effective gauge field within the Chern-Simons
theory, often referred to as the 'hydrodynamic' field, dictates the
behavior of the boundary scalar, which in turn represents the chiral
fermion and expresses its anomalous current
\cite{arouca2022quantum}. It is worth to emphasize that within the
effective bosonic theory, anomalies arise as consequences of the
equations of motion, thereby allowing for a classical description,
with additional considerations required for global effects and
quantization conditions.

The study of topological phases of matter is progressing towards
finding the general topological theories in any dimension, along with
identifying the corresponding anomalies and effective boundary
theories
\cite{Zhang-5d,Cho:2010rk,Putrov-topo,Moy:2022ztf,Frohlich-boso,Palumbo}. This
program has led to considerable advances in understanding
non-perturbative physics. For instance, it has revealed the concept of
fermionic particle-vortex duality in 2+1 dimensions
\cite{Son-duality,Seiberg-duality}, and has elucidated the universal, exact
consequences arising from chiral and gravitational anomalies
\cite{arouca2022quantum,vozmediano-rev}.

We want to stress that the physical setting of topological states,
involving gapped bulk and massless boundary, is ideal for understanding
bosonization in higher dimensions, where topological theories hold
significant importance.

In this work, we combine field theory techniques with the hydrodynamic approach
recently introduced in references
\cite{AbanovI,AbanovII}. Specifically, we describe Euler barotropic
perfect fluids, where local energy depends only on density, using a
variational method based on an action functional. Because viscosity
and heat currents are absent in this hydrodynamic formulation,
it parallels bosonic effective field theory. This connection enables
us to unify results across seemingly different domains of study.

To begin with, we demonstrate a precise correspondence between Euler
hydrodynamics in 1+1 dimensions and the conventional bosonization of
Dirac fermions, encompassing both vector and axial
anomalies. Additionally, we clarify the use of 2+1
dimensional BF topological theory and the inflow mechanism for
identifying the hydrodynamic fields in 1+1 dimensions. It is worth noting that
in this dimension, the axial charge corresponds to the kink
topological charge of the scalar theory.

These results set the stage for the following analysis in 3+1
dimensions. In investigations of anomalous hydrodynamics, the 3+1
dimensional axial charge has been associated with fluid helicity - the
scalar product of vorticity and velocity \cite{Son:2009tf}. We
reformulate and extend the variational formulation of Euler
hydrodynamics and obtain a bosonic `effective field theory'. This
theory accommodates general chiral and mixed axial-gravitational
anomalies of interacting Dirac fermions in fluid phases.

Furthermore, we find the corresponding 4+1 dimensional `bulk'
topological theory which realizes the anomaly inflow and, more
importantly, identifies the needed `boundary' fluid variables.  Beside
the fluid momentum, a new pseudoscalar degree of freedom is
introduced, as a minimum requirement for describing all anomalies of
Dirac fermions, in particular the mixed axial-gravitational anomaly.

The derived 3+1 dimensional hydrodynamics/effective
field theory is applicable to both relativistic and
non-relativistic systems. Importantly, it demonstrates
explicitly that anomalies remain unaffected by the dynamic terms in the
Lagrangian, which specify the hydrodynamic equation of state. This
finding affirms that anomalies encapsulate the system's universal
'geometric' response to background deformations.

The general picture that emerges is that the 4+1d topological theory
serves as a key input for building the effective theory/hydrodynamics.
Additionally, we identify other 4+1 dimensional theories that could
potentially lead to extended hydrodynamics. We discuss one such
theory involving independent vector and axial fluid variables, and
mention another one incorporating two-form fields as degrees of freedom.

The plan of the paper is the following. In Section \ref{sec-2},
we review standard bosonization of 1+1d Dirac fermions and its
anomalies. We present the corresponding 2+1d BF theory originating
from studies of the quantum spin Hall effect, and show how the degrees
of freedom describing bulk and boundary currents can be matched via
the anomaly inflow mechanism. Moving forward to Section
\ref{sec:1+1hydro}, we introduce the action formulation of Euler
hydrodynamics, which is further elaborated on in Appendix
\ref{sec:litvar}. We emphasize its equivalence with the
bosonic theory.

In Section \ref{sec:4} we discuss the hydrodynamics in 3+1
dimensions. We identify the corresponding 4+1 dimensional topological
theory and introduce an additional pseudoscalar field necessary for a
complete description of anomalous currents. In Section \ref{sec:5}, we
incorporate the gravitational background, obtain the spin current, and
discuss the mixed axial-gravitational anomaly. In Section \ref{sec:6},
we discuss the physical interpretation of the new pseudoscalar
field. The Conclusion outlines potential extensions of our
hydrodynamic approach and presents some open questions regarding 3+1
dimensional bosonization.  Appendices \ref{sec:litvar},
\ref{app:coeffs}, and \ref{app:C} respectively cover the variational
formulation of Euler hydrodynamics, the general form of anomaly
coefficients in 3+1 dimensions, and the behavior of fluids in chiral
backgrounds in both 1+1 and 3+1 dimensions.

%%%%%%%%%%%%%%%%%%%%%%%%%%%%%%%%%%%%%%%%%%%%%
\section{1+1 Dimensional anomalies in bosonic field theory
  and anomaly inflow}
\label{sec-2}

%-2.1-%%%%%%%%%%%%%%%%%%%%%%%%%%%%%%%%%%%%%%%%%%%%
\subsection{The scalar theory}
\label{sec-2.1}

We shall start by reminding some known facts about
bosonization in 1+1d, and later discuss
the anomaly inflow from a 2+1d topological theory.
The bosonic action is 
\be
S= -\int d^2x \, (\de_\mu\th)^2\, ,
\label{S-scalar}
\ee
within our conventions for Minkowskian signature
$\eta_{\mu\nu}=diag(-1,1)$.
The theory possesses the $U(1)$ global symmetry $\th\to\th +{\rm const}$,
but two conserved currents
\begin{align}
  J^\mu& =2\de_\mu \th, \qquad\qquad\ \ \ {\rm (Noether\ current)},
 \label{j-top} \\ 
  \tilde J^\mu&=2\eps^{\mu\nu}\de_\nu \th\, , \qquad\qquad
  {\rm (topological\ axial\ current)},
 \label{jt-top}
\end{align}
where hereafter tilded quantities represent pseudo-scalar, pseudo-vectors etc.
Note that the axial current is called topological because it is
conserved without using the equations of motion.
It expresses the kink charge
when a periodic potential is added to the action \eqref{S-scalar}.
As is known, $J, \tilde J$ correspond to vector and axial currents
of Dirac fermions, respectively.

The scalar field can be coupled to both vector and axial backgrounds
$A_\mu, \tilde A_\mu$ as follows:
\be
S= \int d^2x \, -(\de_\mu\th-A_\mu)^2 +
2 \tilde A_\mu\eps^{\mu\nu}(\de_\nu\th-A_\nu)\, ,
\label{11action-100}
\ee
leading to currents
\begin{align}
  J^\mu_{cons} & =\frac{\d S}{\d A_\mu}= 2(\de^\mu\th-A^\mu)
           +2\eps^{\mu\nu}\tilde A_\nu,
  \label{Jcons25}\\
  \tilde J^\mu_{cons} & =\frac{\d S}{\d \tilde A_\mu} =
                        2 \eps^{\mu\nu}(\de_\nu\th -A_\nu),
 \label{Jtcons26}
\end{align}
where the subscript stands for `consistent' current which will be explained
momentarily. The equation of motion obtained by varying (\ref{11action-100})
over $\theta$ is
\be
    \de_\mu(\de^\mu\th-A^\mu) +\de_\mu\eps^{\mu\nu}\tilde A_\nu=0 \, .
 \label{2d-eqm}
\ee
The relations
\begin{align}
  \de_\mu J^\mu_{cons}& =0,
 \label{11vectorD-100}\\
  \de_\mu \tilde J^\mu_{cons} &= - 2\eps^{\a\b}\de_aA_\b \, ,\qquad\qquad
                        \left( \frac{e}{2\pi} \to 1\right),
 \label{11axialD-100}
\end{align}
show that the vector current is conserved (using the equations of motion)
and the axial current is anomalous.
Actually, the form of the action \eqref{11action-100} is manifestly invariant
under the vector gauge transformations
\be
A_\mu(x) \to A_\mu(x) +\de_\mu \l(x)\, ,
\qquad\qquad \th(x) \to\th(x) +\l(x)\, , 
\label{vectransf-100}
\ee
in which the scalar field compensates.
Instead, axial gauge transformations
\be
\tilde A_\mu(x) \to \tilde A_\mu(x) +\de_\mu \tilde \l(x)\, ,
 \label{axtransf-100}
\ee
change the action.
Note that the axial anomaly \eqref{11axialD-100} is normalized
to the Dirac fermion value,
in units of flux quantum, $\Phi_0=e/2\pi$, hereafter set to one.

%-2.2-%%%%%%%%%%%%%%%%%%%%%%%%%%%%%%%%%%%%%%%%%%%%

\subsection{Anomaly inflow, Chern-Simons and
  Wess-Zumino-Witten actions}
\label{sec:2.2}

The 1+1d bosonic theory is assumed to be at the boundary
$\de {\cal M}_3$ of a `bulk'
region  $ {\cal M}_3$ where a gapped topological phase of matter is present.
At energies below the gap, the response to background variations is given
by a topological effective action. The paradigmatic example is given by the
quantum Hall effect (QHE), described by the Chern-Simons (CS) action,
\be
S_{CS}[A]= - \frac{1}{2}\int_{{\cal M}_3}  AdA\, , \qquad\qquad
\left( \frac{e}{2\pi} \to 1\right)\, .
\label{S-CS}
\ee
In this expression, we use the notation of differential forms,
e.g. $A=A_\mu dx^\mu$, omitting the wedge-product symbol $(\wedge)$
for simplicity. It is convenient to write formulas involving
both tensor components and forms; we also represent components of
dual tensors with square brackets, e.g. $[\w_r]^{\mu\nu\cdots}$,
which are defined as follows:
\begin{align}
&  \w_r=\frac{1}{r!}\w_{\mu_1 \dots \mu_r}dx^{\mu_1}\cdots dx^{\mu_r},
  \qquad
  \W_{d-r}=*\w_r= \frac{\sqrt{|g|}}{(d-r)!r!}
  \eps_{\nu_1\dots\nu_{d-r}}^{\ \qquad\mu_1 \dots \mu_r}\, \w_{\mu_1 \dots \mu_r}
  dx^{\nu_1}\cdots dx^{\nu_r},
  \nonumber\\
&  [\w_r]^{\nu_1 \dots \nu_{d-r}} = \frac{1}{r!}
  \eps^{\nu_1\dots\nu_{d-r}\mu_1 \dots \mu_r}\; \w_{\mu_1 \dots \mu_r}\, .
\label{form-def}
\end{align}
Quantities without indices can be easily recognized as being differential forms;
finally, the integral of
the top form is defined as $\int \w_d =\int d^dx\; \w_{1 \dots d}(x)$.

The variation of the Chern-Simons action \eqref{S-CS}
gives the non-dissipative classical Hall
current\footnote{Note that the QHE filling fraction is $\nu=1$,
  corresponding to free fermions on the boundary.}
$j^\mu$ in 2+1d,
\be
j^\mu=\frac{\d S_{CS}}{\d A_\mu}=
-\eps^{\mu\a\b}\de_\a A_\b, \qquad\qquad \mu,\a,\b=1,2,3.
\label{Hcurrent-100}
\ee
\qquad
This bulk current is evaluated in the direction $\hat 3$
orthogonal to the boundary
$\de{\cal M}_3$ located at $x^3=0$, for points $x^\a, \a=1,2$
on the boundary, where it reads
\be
    j^3 =-\eps^{3\a\b}\de_\a A_\b \Big\vert_{\de{\cal M}_3} 
    =\de_\a J^\a, \qquad\qquad \a,\b=1,2;
 \label{Hinflow-100}
\ee
namely, it is equal to the divergence of the 1+1d current $J$. Upon
integration along the one-dimensional space boundary, it implies that
the bulk flux equals the time variation of the charge in the 1+1d
theory. This is the anomaly inflow, the compensation of the quantum
anomaly by a classical current in one extra dimension.

The $\nu=1$ quantum Hall system gives rise to a chiral boundary
fermion. To describe a time-reversal symmetric system, we consider
chiral spin-up and antichiral spin-down fermions at the boundary. The
resulting time-reversal invariant system is realized in the
quantum spin Hall effect and topological insulators
\cite{kane2005quantum,Zhang-rev,arouca2022quantum}.

Neglecting the spin degree of freedom, this setting can describe 1+1d
Dirac fermions. The bulk action is obtained by writing Chern-Simons
actions \eqref{S-CS} for each chiral background $A_\pm=A\pm \tilde A$, that
exchange under time reversal, leading to
\be
S_{BF}[A,\tilde A]= -2\int_{{\cal M}_3} \tilde AdA \,.
 \label{BFaction-100}
\ee
The corresponding bulk currents are
\begin{align}
  j^3&=\frac{\d S_{BF}}{\d A_3} =-2\eps^{3\a\b}\de_\a \tilde A_\b=
       \de_\a J^\b_{cov}\,,
 \label{11covDvec-100} \\
  \tilde j^3&=\frac{\d S_{BF}}{\d \tilde A_3} =-2\eps^{3\a\b}\de_\a A_\b=
              \de_\a \tilde J^\b_{cov}\,.
 \label{11covDax-100}
\end{align}

The 1+1d currents obtained by anomaly inflow $J_{cov}, \tilde J_{cov}$
are called `covariant' and are different from the `consistent'
currents \eqref{Jcons25} \eqref{Jtcons26}, obtained by variation of
the boundary action \eqref{11action-100}, and their respective
anomalies are different.  The two definitions of currents will often
appear in this work: thus, it is important to clarify their technical
and physical properties.\footnote{A full analysis of this subject can
  be found in Sect. 4 and App.~D of \cite{arouca2022quantum}.}
Covariant and consistent currents are best known in non-Abelian gauge
theories \cite{bardeen1984consistent}, but they also occur in presence
of both Abelian backgrounds $A,\tilde A$, where `covariant' actually
stands for invariant.

The 2+1 bulk currents $j, \tilde j$ are manifestly gauge invariant,
and their reduction in 1+1d, $J_{cov},\tilde J_{cov}$ are also
invariant, as explicitly shown in the following Section. Gauge
invariant currents should be used when coupling to any physical
quantity, either in 1+1d or 2+1d. The 2+1 bulk here plays the role of a
charge reservoir for the 1+1 boundary. It is known that
covariant anomalies can be obtained by
heat-kernel regularization of the fermion path-integral, they obey
the Index theorem and describe the spectral flow.

Consistent currents $J_{cons},\tilde J_{cons}$ are not gauge invariant
in general; they are obtained by a variation of the 1+1d effective
action, as seen before, or by fermion loop calculations.  They are
also derived from the Wess-Zumino-Witten (WZW) action, obeying the
Wess-Zumino consistency conditions, thus explaining their name
\cite{wess1971consequences}. These currents describe the physics of
isolated 1+1d systems, with some provisos.

The relation between the Chern-Simons and  WZW action,
\be
S_{CS}[A+d\l,\tilde A+d\tilde\l]=S_{CS}[A,\tilde A] +
S_{WZW}[A,\l,\tilde A,\tilde\l]\, ,
 \label{WZ-action-100}
\ee
shows some properties of the two kinds of currents.
Inserting the action \eqref{BFaction-100}, we obtain from
(\ref{WZ-action-100})\footnote{
  The WZW action is 1+1 dimensional in the Abelian case.}
\be
S_{WZW}=- 2\int_{\de {\cal M}_3} \tilde \l dA\, ,
\label{S-WZW}
\ee
whose variations over $\lambda,\tilde\lambda$ correctly
reproduce the consistent anomalies
(\ref{11vectorD-100},\ref{11axialD-100}).

Note that another form of the bulk action \eqref{BFaction-100},
$S_{BF}=-2\int Ad\tilde A$,
differing by a boundary term, would give the same
covariant currents \eqref{11covDvec-100},\eqref{11covDax-100},
but would lead to different WZW action and consistent anomalies,
\be
    \hat S_{WZW}=-2\int_{\de {\cal M}_3} \l d\tilde A, \qquad\qquad 
    \de_\mu \hat J^\mu_{cons}=-2\eps^{\a\b}\de_\a \tilde A_\b\,, 
    \qquad\qquad
    \de_\mu \hat {\tilde J}^\mu_{cons}=0\,.
\label{11vecaxD-150}
\ee
The difference between $J_{cons}$ and $\hat J_{cons}$ (and their axial
companions) is related to the freedom of changing the 1+1d effective
action by terms polynomial in the backgrounds. This can be used,
e.g. to switch anomaly from one symmetry to the other. However, the
relation between $J_{cons}$ and $J_{cov}$ is not of this kind. The
differences,
\be
    J_{cov}=J_{cons}+\D J, \qquad\qquad 
    \tilde J_{cov}=\tilde J_{cons}+\D \tilde J,
\label{BZterms-100}
\ee
are called Bardeen-Zumino terms \cite{bardeen1984consistent};
they can be computed by paying attention to boundary terms when varying the
bulk action \eqref{BFaction-100}. We find
\be
    \d S_{BF} = -2\int_{{\cal M}_3} \d \tilde A dA +\d A d\tilde A \ -\ 
    2 \int _{\de {\cal M}_3} \d A \tilde A\, .
 \label{varBF-100}
\ee
From the variation of the bulk term we recover the covariant currents
\eqref{11covDvec-100},\eqref{11covDax-100}, while from the boundary
term we get the Bardeen-Zumino
terms
\be
    \D J^\a=-2 \eps^{\a\b} \tilde A^\b, \qquad\qquad \D \tilde J^\a=0\, ,
 \label{dif-100}
\ee
which match the differences between the two kinds of anomalies
\eqref{11vectorD-100}\eqref{11axialD-100} and \eqref{11covDvec-100},
\eqref{11covDax-100}.

In summary, the  covariant currents are obtained by anomaly
inflow and appear in gauge invariant expressions; the
consistent currents follow by variation of the 1+1d effective
action. The relation between the two kinds of currents can be
unambiguously determined from the 2+1d topological theory \eqref{BFaction-100}.

%-2.3-%%%%%%%%%%%%%%%%%%%%%%%%
\subsection{`Hydrodynamic' topological theory and bosonic fields}
\label{sec:23}

Topological phases of matter are characterized by global effects and
excitations which are described by topological actions written in
terms of matter currents. Borrowing the reasoning from hydrodynamics,
after integrating out high-energy excitations, the low-energy states
should be described in terms of conserved currents \cite{Liu:2018kfw}.
Furthermore, these currents can be parameterized by dual gauge fields:
in 2+1d we can write
\be
    j^\mu=\eps^{\mu\nu\r}\de_\nu \tilde q_\r\,,
    \qquad\qquad
    \tilde j^\mu=\eps^{\mu\nu\r}\de_\nu p_\r\,,
 \label{j-3d}
\ee
where $p, \tilde q$ are one-form dynamic gauge fields expressing
matter degrees of freedom. The parameterization (\ref{j-3d}) makes
conservation of currents automatic (`topological currents').

Regarding the interaction between these fields, we can introduce a BF term
leading to the following `hydrodynamic' topological theory,
\be
    S[p,\tilde q; A,\tilde A]=2\int_{{\cal M}_3} \tilde q dp 
    + \tilde q d A +  p d\tilde A \,,
 \label{CS}
\ee
which has been used for describing the quantum spin Hall effect
\cite{arouca2022quantum}.

Upon integration in $p, \tilde q$, one correctly obtains the
`response' action \eqref{BFaction-100} $S_{BF}[A,\tilde A]$ involving
the backgrounds only. The action \eqref{CS} for general coupling
constant $2k$ (here $k=1$)
describes anyonic braiding of excitations coupled to the
$p,\tilde q$ fields and the ground state degeneracy on topologically
non-trivial manifolds, the so-called topological order
\cite{Wenbook}.

In the following we show how the bulk $p,\tilde q$ fields are related
to the boundary degrees of freedom which define
the scalar theory of Section \ref{sec-2.1}. This identification follows
by matching the covariant currents in the anomaly inflow.
We first solve the equations of motion of the bulk theory \eqref{CS},
with the result
\be
dp+dA=0 \ \ \to\ \    p=d\th -A , \qquad\qquad
d\tilde q +d\tilde A=0 \ \ \to\ \ \tilde q=d\psi -\tilde A,
 \label{eomsolv-100}
\ee
where $\th, \psi$ are respectively scalar and pseudoscalar
gauge degrees of freedom, which are free variables at this level. Then, we
reconsider the inflow relation \eqref{11covDax-100}
\be
    \tilde j^3=2\eps^{3\a\b}\de_\a p_\b =\de_\a \tilde J^\a_{cov}
    \quad \to \quad
    \tilde J^\a_{cov} =2\eps^{\a\b}p_\b \underset{eq.m.}{=}
    2 \eps^{\a\b}(\de_\b\th - A_\b)\, .
 \label{jtcov-2d}
\ee
In the identification of the 1+1d current $\tilde J_{cov}$,
we used the equations of motion and the freedom in choosing $\th$.
An analogous expression is found for the current $J_{cov}$,
\be
    J^\a_{cov}=2\eps^{\a\b}\tilde q_\b
    \underset{eq.m.}{=}2\eps^{\a\b}(\de_\b\psi - \tilde A_\b)\, ,
 \label{jcov-2d}
\ee
in terms of the pseudoscalar field $\psi$.

We now remark that 2+1d currents are manifestly gauge invariant,
thus the 1+1d covariant currents should also be so. This requirement
implies a boundary gauge condition for the bulk fields, namely:
\be
    p=d\th -A\; ,\qquad \tilde q=d\psi -\tilde A\, ,
    \qquad\qquad {\rm gauge\ invariant\ on} \ \de{\cal M}_3\,.
 \label{g-cond}
\ee
It follows that the fields $\th,\psi$ must compensate the
gauge transformations of background fields:
$(A\to A+d\l, \th\to \th+\l)$ and
$(\tilde A\to\tilde A+d\tilde\l, \psi\to \psi+\tilde\l)$.

We see that the inflow relation \eqref{jtcov-2d} reproduces the axial current
\eqref{Jtcons26} as a function of previously introduced scalar field
$\th$ with correct gauge transformations.  Indeed, this field is the
gauge degree of freedom that becomes physical, as it occurs in
$\s$-models of spontaneous symmetry breaking and the Higgs phenomenon;
being a consequence of gauge symmetry only, the same mechanism also
occurs in other phases of matter, having different dynamics.

The complete form of the effective action including bulk and boundary
terms can be written as follows\footnote{
  In this and following integrals we use a mixed notation
  involving both vectors (with indices) and
  differential forms (without indices), with rules specified earlier in
  \eqref{form-def}.}
\be
    S= \int_{{\cal M}_2} - p_\mu^2  +
    2\int_{{\cal M}_3} \tilde q dp+\tilde q dA +pd\tilde A\, .
 \label{2+1d-act}
\ee

In this expression, the first term encodes the dynamics of boundary
excitations, while the second term is the topological action
(\ref{CS}).

It is also convenient to introduce the notations
$\pi=p+A$ and $\tilde\pi=\tilde q+A$ and rewrite (\ref{2+1d-act}) as
\begin{align}
   S= &\int_{{\cal M}_2} - (\pi_\mu-A_\mu)^2  + 2\tilde A(\pi-A) +
                2\int_{{\cal M}_3} \tilde \pi d\pi-\tilde AdA\, ,
                \label{2+1d-act-pi}\\
&\qquad  \pi=p+A, \qquad\qquad \tilde\pi=\tilde q+A\,,
  \label{pi-tildepi-def}
\end{align}
paying attention to boundary terms.

Upon using the bulk equations of motion \eqref{eomsolv-100}
for $p,\tilde q$,  which now read $\pi=d\theta$ and $\tilde \pi=d\psi$,
one is left with an action which depends on the yet
undetermined 1+1d fields $\th,\psi$ at the boundary, as follows
\be
    S=\int_{{\cal M}_2} -(\de_\mu\th -A_\mu)^2
    + 2 \tilde A\left(d\th -A \right)
    \ - 2\int_{{\cal M}_3} \tilde A dA\,.
 \label{th-act}
\ee

The expressions \eqref{th-act} and \eqref{2+1d-act-pi}
of combined bulk-boundary theory are manifestly
gauge-invariant, according to \eqref{g-cond}: the 1+1d part reproduces
the bosonic theory \eqref{11action-100} described in Section \ref{sec-2.1}.
Thus, we succeeded in re-deriving the boundary bosonic theory from bulk
topological data and anomaly inflow.

Let us add some remarks:
\begin{itemize}
\item The form of the bulk-boundary action
  \eqref{2+1d-act-pi} is uniquely obtained by gauge invariance from
  the topological theory \eqref{CS}, apart from the addition of
  the gauge invariant dynamic term $p_\mu^2$.
  The ambiguities on boundary terms which might occur in defining the
  2+1d theory \eqref{CS} have been fixed by earlier assumptions
  \eqref{11vecaxD-150}, \eqref{BFaction-100}, determining the form
  of consistent currents on the boundary.

\item Bulk and boundary variables have been matched by comparing
  currents via inflow \eqref{jtcov-2d}. \eqref{jcov-2d}.
  These observable quantities are free of ambiguities.
  The boundary degrees of freedom can also be obtained by other methods,
  but we find that the matching of currents provides the most
  reliable procedure. Note that $\psi$ has disappeared from \eqref{th-act}
  but is relevant for other forms of the action, as shown in the next
  Section.
 
\item The bulk equations of motion \eqref{eomsolv-100} for
  $p,\tilde q$ were obtained by assuming vanishing variations on the
  boundary.  Their solution is extended by continuity to the
  boundary. Next, the remaining undetermined degrees of freedom
  $\th,\psi$ are subjected to the boundary equations of motion and
  actually acquire a dynamics due to the first term in
  \eqref{th-act}. 

\item
  Finally, a general remark on using classical actions: they may have
  some arbitrariness because different functional forms can share the same
  extrema. For example, it is possible to solve a subset of
  equations of motion and substitute them into the action,
  obtaining another, equivalent functional, to be extremized on the remaining
  variables.
\end{itemize}

%-2.3.1-%%%%%%%%%%%%%%%%%%%%%%%%%%%%%%%%%%%%%%%%%%%%

%-------------------------------------------------------
\subsubsection{Duality of 1+1 dimensional theory}

The scalar fields $\psi,\th$ play a rather symmetric role in
the 2+1d action \eqref{CS} and covariant currents \eqref{jtcov-2d},
\eqref{jcov-2d}: actually, the theory
is manifestly invariant under the following
duality that exchanges fields, currents and anomalies,
\be
(\th, J_{cov}, A) \leftrightarrow (\psi, \tilde J_{cov},\tilde A).
\label{duality-100}
\ee

The boundary theory \eqref{th-act} is not explicitly self-dual.
The field $\psi$ cancels out and the consistent currents \eqref{Jcons25},
\eqref{Jtcons26} are of different nature:
$\tilde J_{cov}= \tilde J_{cons}$ is topological, while
$ J_{cons}$ is of Noether type, obeying $\de_\mu J^\mu_{cons}=0$,
Actually, the duality is also present in 1+1d: it
is expressed by mapping the action \eqref{th-act} into another one with the
same functional from in terms of dual variables.

We start again from the bulk-boundary expressions \eqref{jtcov-2d},
\eqref{jcov-2d}, and substitute the following partial solution of the
bulk equations of motion: $p=\pi -A$ and $\tilde q=d\psi-\tilde A$,
where $\pi$ transforming as $A$ for respecting the
gauge condition \eqref{g-cond}.

The action \eqref{2+1d-act-pi} becomes (neglecting the inert background part
$\int\!\!\tilde A dA$)
\be
\tilde S=2\int_{{\cal M}_2} - \frac{1}{2}(\pi_\mu -A_\mu)^2 +
\tilde A\left(\pi -A \right) +\psi d\pi \, .
\label{rel-hydro}
\ee
The pseudoscalar $\psi$ is now present and it enforces the constraint
$d\pi=0$. Solving it by $\pi=d\th$, the 
action \eqref{th-act} is reobtained. Thus, \eqref{rel-hydro} is
an equivalent form of \eqref{th-act}.

The equations of motion can be equivalently obtained by varying with
respect to $\pi$, leading to
\be
    \pi^\mu -A^\mu=\eps^{\mu\nu}(\de_\nu \psi -\tilde A_\nu )\, .
 \label{eom-233}
\ee
This equation can be used to eliminate $\pi$ in \eqref{rel-hydro}, with the
result
\be
\tilde S=2 \int_{{\cal M}_2} -\frac{1}{2}(\de_\mu\psi -\tilde A_\mu)^2 +
A\left(d\psi -\tilde A \right) +A \tilde A \, .
 \label{th-act-dual}
\ee
This is the dual theory of \eqref{th-act}, having the same form
under the exchanges \eqref{duality-100}.
It is expressed in terms of the
pseudoscalar field and the consistent currents \eqref{Jcons25},
\eqref{Jtcons26}, exchange their nature:
$\tilde J$ is of Noether type, obeying $\de_\nu\tilde J^\nu_{cons}=0$, and
$J_{cons}=J_{cov}$ is topological.

The equations of motion \eqref{eom-233} for $A=\tilde A=0$ and
$\pi=d\th$ become
\be
\de^\mu\th = \eps^{\mu\nu} \de_\nu\psi\, ,
\label{2d-duality}
\ee
showing that $\psi$ is the dual field of $\th$.
In conclusion, the covariant currents obtained by inflow \eqref{jcov-2d},
\eqref{jtcov-2d}
are both of topological kind, with $\psi$ the dual field of $\th$.
The consistent currents are expressed in terms of a single field,
one is topological and the other of Noether type.

In the Appendix \ref{app:C} we consider the bosonic theory \eqref{th-act}
in a chiral background, i.e. $A=\tilde A$ and
show that boundary excitation split into chiral components, one
interacting with the background and the other one remaining inert.

Let us finally recall the description of interacting fermions
by the bosonic theory, which is the main virtue of 1+1d
bosonization.\footnote{
A complete account can be found in \cite{Ginsparg,arouca2022quantum}.}
The analysis in previous sections has shown how the boson
maps into the free fermion and reproduces its anomalies.
The bosonic theory admits a simple generalization. Starting from the
topological action \eqref{CS}, a coupling constant $k$ can be added 
to the first term, $2 \int \tilde q d p \ \to\  2 k\int \tilde q d p$,
leading to anomalies \eqref{jtcov-2d},\eqref{jcov-2d}
multiplied by the factor $1/k$, as well as the corresponding WZW action.
As a consequence, the 1+1d bosonic theory \eqref{th-act} is also
modified by the coupling $k$, which parameterizes a critical line
describing massless fermions with current-current interaction.
The duality transformation of the previous section also extends
to a map between theories with $k \leftrightarrow 1/k$.

At the quantum level, the bosonic field becomes a compact variable,
$\th(x) \equiv \th(x) +2\pi$, for the proper definition of large gauge
transformations on physical states (cf. \eqref{g-cond}).
The coupling constant $k$ can be included in $\th$ by a field
redefinition, thus changing the compactification radius: within this
convention, the critical line is equivalently parameterized
by the  compactification radius.

%-3-%%%%%%%%%%%%%%%%%%%%%%%%%%%%%%%%%%%%%%%%%%%%
\section{Hydrodynamics with anomalies}
 \label{sec:1+1hydro}

%-3.1-%%%%%%%%%%%%%%%%%%%%%%%%%%%%%%%%%%%%%%%%%%%%
\subsection{Variational formulation of Euler hydrodynamics}

In this Section we recall the description of 
perfect barotropic fluids (in any dimension) developed in the
Refs.\cite{AbanovI}\cite{AbanovII}.
In these systems, the temperature vanishes, entropy is constant and there is
no heat conduction, viscosity and shear. Furthermore,
the pressure depends on density only.
The subject has a very long history and some introductory
material can be found in Appendix \ref{sec:litvar} and the
review \cite{jackiw-rev}.
The same formulation applies to both non-relativistic
and relativistic fluids: the first case is briefly discussed here
for simplicity.\footnote{
  The relativistic case is discussed in the Appendix \ref{sec:litvar},}

The Lagrangian variational problem is obtained starting from
the Hamiltonian description of the Euler equation in terms of the
basic variables of density $\r$ and velocity $v^i$, obeying Poisson brackets.
The `particle current' is defined as 
${\cal J}^\mu=(\r, \r v^i)$ and satisfies $\de_\mu{\cal J}^\mu$=0.
It is possible to Legendre transform from energy $\eps(\r)$ 
to pressure $P(\mu)$, which is a function of the chemical potential $\mu$,
to obtain the action,
\be
S= \int \r \mu -\eps(\r) =\int P(\mu), \qquad\qquad \frac{dP}{d\mu}=\r,
\quad\frac{d\eps}{d\r}=\mu\, .
\ee

The chemical potential itself depends on the particle momenta,
$\mu=\mu(p_\a)$, with dependence specific for nonrelativistic
(Galilean) and relativistic (Lorentzian) fluids (See Appendix \ref{sec:litvar}). 
Thus,  the variational approach is defined for 
the following functional of momenta $p_\a$
\be
S[p_\a]=\int P(p_\a)\, .
\label{act32}
\ee

This formulation of hydrodynamics is 
very similar to an effective field theory for the `field' $p_\a$.
Both are low-energy descriptions, and we are not dealing with entropy
and dissipation, which would require more advanced quantum field theory
settings, such as the Schwinger-Keldysh approach
\cite{Liu:2018kfw,Jensen:2018hse}.

However, there is still a difference, because the variational problem
for the hydrodynamic action \eqref{act32} is constrained. As explained
in  the review \cite{jackiw-rev}, in going from the Lagrangian
fluid description\footnote{Not to be confused with the Lagrangian
  functional.} to the Eulerian formulation, the motion of each fluid
element is lost, leaving a reduced set of degrees of freedom that obey
constraints. Due to this reduction, the Hamiltonian
Poisson algebra obeyed by $(\r, v^i)$ possesses extra
Casimirs which prevents the writing of a standard Lagrangian
variational principle (see also Appendix \ref{sec:litvar}).

As a consequence, the variations $\d p_\a$ 
extremizing \eqref{act32} are constrained: one possibility is to use a reduced
functional dependence for $p_\a$ given by the Clebsch parameterization
\cite{jackiw-rev}.  Another formulation uses restricted,
`admissible' variations \cite{carter1988standard}
which correspond to the diffeomorphisms, expressed by the Lie derivative
$\d_\eps p={\cal L}_\eps p$
\be
\d S[p]=0\, , \qquad {\rm for} \qquad
\d_\eps p_\nu=\varepsilon^\mu\p_\mu p_\nu + p_\mu\p_\nu\varepsilon^\mu,
\label{Lie-der}
\ee
where $\eps^\mu$ is an arbitrary vector field.

The variation of the action \eqref{act32} leads to the following two results: 

{\bf I}. We identify the particle current ${\cal J}$, and obtain
after one integration by parts
\be
{\cal J}^\mu=-\frac{\d S}{\d p_\mu}= -\frac{\de P}{\de p_\mu}, \qquad\qquad
\mathcal{J}^\nu (\p_\mu p_\nu-\p_\nu p_\mu)+p_\nu \de_\mu \mathcal{J}^\mu=0
\, .
\ee
Upon contracting  the second relation with ${\cal J}^\mu$,
we get two equations of motion
\be
\de_\mu \mathcal{J}^\mu=0, \qquad\qquad
\mathcal{J}^\nu (\p_\mu p_\nu-\p_\nu p_\mu)=0,
\label{CL35}
\ee
corresponding to current conservation and 
the so-called Carter-Lichnerowicz (CL) equation
\cite{lichnerowicz1994relativistic}.
As shown in Appendix \ref{sec:litvar}, the latter is equal to the Euler equation after
expressing the variables ${\cal J}^\mu,p_\mu$ in terms of $\rho,v^i$ and the
chemical potential $\mu$.

{\bf II}. Given that the Lie derivative expresses reparameterization
invariance, the variation \eqref{Lie-der} defines the stress tensor of the fluid
and establishes its conservation: we find,
\be
    T_\mu^\nu = -\frac{\p P}{\p p_\nu}p_\mu +\delta^\nu_\mu P\,,
    \qquad\qquad \de_\nu T_\mu^\nu=0.
 \label{t-hydro}
\ee
These equations reproduce again the Euler equation (for spatial $\mu$)
and the energy conservation (for $\mu$ the time direction).
The relativistic fluid dynamics is also obtained 
for specific forms of pressure $P(p_\a)$ and variables $p_\mu$, as
shown in Appendix \ref{sec:litvar}.

% -3.1.1-%%%%%%%%%%%%%%%%%%%%%%%%%%%%%%%%%%%%%%%%%%%%

\subsubsection{Topological currents in 1+1 and 3+1 dimensions}
\label{sec:311}

The Carter-Lichnerowicz equation \eqref{CL35} can be
rewritten in differential-form notation using the inner product $i_X$,
\be
    i_{\cal J} dp=0, \qquad\qquad {\cal J}^\mu=-\frac{\de P}{\de p_\mu}\, .
 \label{CL-37}
\ee
Note that this is not a diffeomorphism-invariant equation, because the
current ${\cal J}$ depends itself on $p_\a$ and contains dynamic
information through the form of pressure $P$ .

However, even spacetime dimensions are special. In 1+1d we observe that $dp$ is
a top form, dual to a scalar quantity: the CL equation is
actually a scalar condition. Thus, if this is satisfied for
the particle current ${\cal J}$, it also holds for arbitrary currents:
it amounts to the constraint
\be
    dp=0\, , \qquad\qquad\qquad\qquad {\rm (1+1\ d)}.
\label{dpi-0}
\ee
 In 3+1d, by multiplying the CL equation by $dp$,
it gives the condition $i_{\cal J} (dpdp)=0$ for a top form,
which similarly implies the constraint $dpdp=0$.

Therefore, in even spacetime dimensions the CL equation results in a
geometric condition independent of dynamics.
It leads to the definition of conserved topological currents, as follows
\begin{align}
  1+1d: &\qquad dp=0 \qquad \to\ \qquad \de_\mu \tilde J^\mu=0\,,
          \qquad\qquad \tilde J^\mu=\eps^{\mu\nu}p_\nu\, ,
          \\
  3+1d: &\qquad dpdp=0 \quad \to\ \qquad \de_\mu \tilde J^\mu=0\,,
\qquad\qquad \tilde J^\mu=\eps^{\mu\nu\r\s}p_\nu\de_\r p_\s\, .
\label{top-curr}
\end{align}

In 1+1d, one recovers the axial current \eqref{jt-top} in the bosonic form;
in 3+1d a new axial current is conserved by the flow, whose charge is
the so-called fluid helicity
\be
    \tilde Q=\int d^3x\ \tilde J^0 \propto \int d^3x\, \vec v \cdot\vec \w,
    \qquad \vec \w =\nabla \times \vec v\, ,
 \label{Qt}
\ee
which is the integral of the scalar product of vorticity and velocity.

In conclusion, the hydrodynamic approach in even spacetime dimension
possesses both vector and axial currents. This rather remarkable fact
allows for the description of fermionic systems and their anomalies,
which will be analyzed in detail in the following discussion.

%-3.2-%%%%%%%%%%%%%%%%%%%%%%%%%%%%%%%%%%%%%%%%%%%%

\subsection{Equivalence of hydrodynamics and bosonization
  in 1+1 dimensions}

\subsubsection{Admissible variations}

The coupling to backgrounds $A, \tilde A$ is introduced
in the hydrodynamic action as follows. The kinetic momentum is
replaced by the canonical momentum $\pi$,
\be
p=\pi -A\, ,
\label{pi-def}
\ee
where $\pi$ varies as $A$ under gauge transformations for keeping $p$
gauge invariant (cf. \eqref{g-cond}).
Next, $\tilde A$ is coupled to the axial current
found in the previous Section. The action becomes
\begin{align}
    S[\pi] &= \int d^2x\, P(\pi-A) +2\int_{{\cal M}_2} \tilde{A}(\pi-A)\,.
 \label{action-100}
\end{align}

As discussed in the previous Section, the variational problem is
given by diffeomorphism variations $\d \pi={\cal L}_\eps \pi$,
and the result is the CL equation, implying a constraint
\be
    \mathcal{J}^\nu (\p_\mu\pi_\nu-\p_\nu\pi_\mu) = 0\,,
    \qquad \rightarrow\qquad d\pi = 0\,,
 \label{dpi-100}
\ee
together with the conservation of the particle current
\be
\de_\nu \mathcal{J}^\nu=0, \qquad
\mathcal{J}^\nu = -\frac{\p P}{\p \pi_\nu} +
2\epsilon^{\nu\mu}\tilde A_\mu\,,
\label{CL-2d}
\ee
which now includes a background term.

The consistent currents are obtained by variation of the action
\begin{align}
  J^\nu_{cons} &= \frac{\delta S}{\delta A_\nu} =
                 -\frac{\p P}{\p \pi_\nu} +2\eps^{\nu\mu}\tilde A_\mu
                 = {\cal J}^\nu\, ,
 \\
    \tilde J^\nu_{cons} &= \frac{\delta S}{\delta \tilde A_\nu} =
                          2\eps^{\nu\mu}(\pi_\mu-A_\mu)\, .
                          \label{j-2d}
\end{align}
Upon using the equations of motion, their divergences
reproduce the expected expressions of Section \ref{sec-2.1},
\be
    \de_\nu J^\nu_{cons}=0, \qquad\qquad\qquad
    \de_\nu \tilde J^\nu_{cons}=-2 \eps^{\nu\mu}\de_\nu A_\mu\, .
 \label{anom-2d}
\ee
We conclude that the hydrodynamic theory is able to describe Dirac
fluids in 1+1 dimensions.

The conservation law for the stress tensor is obtained by applying a
diffeomorphism to all fields in the action \eqref{action-100}:
\begin{align}
  \delta S &= \int d^4x\, \left(\frac{\delta S}{\delta \pi_\nu}\d_\eps\pi_\nu+
             \frac{\delta S}{\delta A_\nu}\delta_\eps A_\nu +
  \frac{\delta S}{\delta \tilde A_\nu}\delta_\eps \tilde A_\nu \right)
\nonumber \\
&= \int d^4x\, \epsilon^\mu\left( - \de_\nu T^\nu_\mu+J_{cons}^\nu \p_\mu A_\nu -
      \p_\nu(A_\mu J^\nu_{cons}) + \tilde J^\nu_{cons} \p_\mu \tilde A_\nu -
      \p_\nu(\tilde A_\mu \tilde J^\nu_{cons}) \right)
\nonumber  \\
&= \int d^4x\, \epsilon^\mu\left( - \de_\nu T^\nu_\mu+F_{\mu\nu}J^\nu_{cons} -
             A_\mu\p_\nu J^\nu_{cons} +\tilde F_{\mu\nu}\tilde J^\nu_{cons} -
             \tilde A_\mu\p_\nu \tilde J^\nu_{cons}\right)\, .
 \label{deltaS-310}
\end{align}
In these equations, all variations are Lie derivatives \eqref{Lie-der} with
parameter $\eps$, the first term giving the stress tensor divergence,
with
\be
    T_\mu^\nu = -\frac{\p P}{\p \pi_\nu}(\pi_\mu-A_\mu) +\delta^\nu_\mu P\,.
 \label{T-def}
 \ee
After rewriting the consistent currents in terms of covariant
 ones, one obtains\footnote{The result can also be checked from
   definitions of $T,J,\tilde J$, using the equation of motion
   \eqref{CL-2d} and the 1+1d identity $F_{\mu\nu}=F\eps_{\mu\nu}$,
   $F=\de_1A_2-\de_2A_1$.}
\cite{AbanovII}
\be
  \p_\nu T_\mu^\nu = F_{\mu\nu}J^\nu_{cov} +
  \tilde F_{\mu\nu}\tilde J^\nu_{cov}\, .
  \label{T-cons}
\ee
Note that the currents appearing in the stress tensor equation are
of the covariant type, in agreement with gauge invariance.

The geometric condition $d\pi=0$ can be solved in terms of a scalar
variable $\pi=d\th$; upon this substitution, it is clear that the
hydrodynamic action and the currents exactly corresponds to expressions
of the bosonic theory of Section \ref{sec-2.1}. In this correspondence,
the relativistic expression of the pressure
$P(p_\a)\sim p_\a^2$ maps in the dynamic term of the bosonic action
$P(d\th -A) \leftrightarrow - (\de_\mu\th -A_\mu)^2$.
Actually, in the hydrodynamic as well as the bosonic theory, one can
check that the anomalies are completely independent of the dynamics,
provided that gauge invariance is respected. In particular, they also
occur for non-relativistic and vanishing dynamics ($P=0$).

Of course, it is well known that anomalies are geometric, i.e.
largely independent
of interactions, but the present approach makes this property
manifest. This is one interesting aspect of it.

%-3.2.2-%%%%%%%%%%%%%%%%%%%%%%%%%%%%%%%%%%%%%%%%%%%%

%--------------------------------------------------
\subsubsection{Free variations with explicit constraint}
\label{sec:322}

One difference remains to be understood before establishing the complete
correspondence between  bosonic field theory and hydrodynamics in
1+1d: the field-theory Lagrangian is extremized under general 
variations, while they are restricted in hydrodynamics.

Recall that admissible variations \eqref{Lie-der} imply the simple constraint
$d\pi=0$ \eqref{dpi-100} in 1+1d.  Thus, we can freely vary $\d \pi$ by
adding the constraint via a (pseudoscalar) Lagrange multiplier $\psi$.

The action \eqref{action-100} becomes
\begin{align}
  S[\pi,\psi] &= \int d^2x\, P(\pi-A) +2 \int_{{\cal M}_2} \tilde{A} (\pi-A) +
            \pi d\psi\,.
 \label{2d-constr}
\end{align}
The expressions of the stress tensor \eqref{T-def} and currents \eqref{j-2d}
are clearly unchanged. The equations of motion for
the unconstrained variables $\pi$ and $\psi$ are
\begin{align}
 \psi:& \qquad d\pi=0 \qquad\to\qquad \pi_\mu=\de_\mu\th  \,,
\label{new-em}
\\
  \pi:& \qquad\frac{\de P}{\de\pi_\mu}=
        2\eps^{\mu\nu}\left( \tilde A_\nu-\de_\nu\psi\right) \,.
\label{new-em2}
\end{align}
Using them in the (unchanged) expressions of the currents \eqref{j-2d}
and stress tensor \eqref{T-def},
one  verifies that the same anomalies and conservation equations
\eqref{anom-2d}, \eqref{T-cons} are reobtained.

More precisely, the new equations of motion \eqref{new-em}, \eqref{new-em2}
can be compared
with those obtained by admissible variations \eqref{dpi-100}, \eqref{CL-2d}:
\begin{align}
    {\rm admissible:}\qquad &d\pi =0,\qquad\quad
    \de_\mu J_{cons}^\mu=
    \de_\mu\left(-\frac{\de P}{\de \pi_\mu} +2\eps^{\mu\nu}\tilde A_\nu\right),
 \\
    {\rm free:}\qquad\qquad\quad  &d\pi=0,  \qquad\qquad
    J_{cons}^\mu=\left(-\frac{\de P}{\de \pi_\mu} +2\eps^{\mu\nu}\tilde A_\nu\right)
      \underset{eq.m.}{=}2\eps^{\mu\nu}\de_\nu\psi \,.              
 \label{new-em3}
\end{align}
It is apparent that current conservation has been replaced by its
`first integral', namely the general parameterization of a conserved current
in terms of the unconstrained parameter $\psi$.
Note that in classical mechanics, freely varying the action with
a time-dependent constraint imposed by a Lagrange multiplier
is generally not equivalent to taking restricted variations.
However, in the present case we checked that the same result is obtained.

In conclusion, we have shown that the hydrodynamic theory can
be reformulated as the unconstrained Lagrangian variational problem
\eqref{2d-constr}, and as such it is completely equivalent to standard
field-theory bosonization in 1+1 dimensions.

%-3.3-%%%%%%%%%%%%%%%%%%%%%%%%%%%%%%%%
\subsection{Summary of Sections two and three}

In Section \ref{sec-2} we recalled some properties of bosonization in 1+1d and
got acquainted with different forms of currents and their relation with
Chern-Simons and WZW actions. We then showed that the bosonic theory can
be inferred from the `hydrodynamic' topological theory in one extra dimension,
which not only determines the anomalies but also suggests the needed
boundary degrees of freedom.

In Section \ref{sec:1+1hydro}, we reviewed the hydrodynamic approach
developed in Refs.\cite{AbanovI}, relying on an action functional. We
observed that the equations of motion imply dynamics-independent
constraints in 1+1 and 3+1d, thus allowing free
variational problems as those considered in quantum field theory. In
1+1d we established the complete correspondence between this
formulation of hydrodynamics and the well-known bosonic field theory
describing fermions.

Based on this background material, in the next Section we are going to
develop the hydrodynamics with anomalies in 3+1d, extending the
results of \cite{AbanovII}. The input from the `hydrodynamic' topological
theory in 4+1d will suggest new variables which are necessary for a
complete description of anomalous fluids in 3+1 dimensions.

%-4-%%%%%%%%%%%%%%%%%%%%%%%%%%%%%%%%%%%%%%%%%%%%

\section{Hydrodynamics and inflow in 3+1 dimensions }
\label{sec:4}

%-4.1-%%%%%%%%%%%%%%%%%%%%%%%%%%%%%%%%%%%%%%%%%%%%

\subsection{4+1 Dimensional Chern-Simons action and
  general form of anomalies}
\label{sec:red4+1top}

The `bulk' response action is well known in the literature
\cite{bertlmann2000anomalies,arouca2022quantum}: it is given by the
4+1d Chern-Simons action written in terms of background fields.

We start from the expression describing
the chiral anomaly of a 3+1d Weyl fermion\footnote{
  See Appendix \ref{app:coeffs} for the normalization of fields.}
\be
S[A_+]= -\frac{1}{6}\int_{{\cal M}_5} A_+ d A_+ d A_+\,, \qquad\qquad
\left(\frac{e}{2\pi}\to 1 \right)\,,
\label{S-weyl-resp}
\ee
where the chiral decomposition is defined by
\be
A_\pm=A\pm \tilde A,\qquad J_\pm=\frac{1}{2}(J\pm \tilde J) \qquad\to\qquad
J^\mu A_\mu +\tilde J^\mu \tilde A_\mu=
J_+^\mu A_{+\mu} +\tilde J_-^\mu \tilde A_{-\mu}\,.
\label{chi-split}
\ee
Note that this action is time reversal invariant but not
parity invariant, as it should.
Then one subtracts to it another copy $S[A_-]$, thus obtaining
the parity invariant expression describing Dirac fermions,
\be
S[A, \tilde A]= -\int_{{\cal M}_5} \tilde A d A d  A
+\a \tilde A d \tilde A d \tilde A\,, \qquad\qquad \a=\frac{1}{3}\,,
\label{S5-resp}
\ee
with $A_\pm=A \pm \tilde A$.
The action  \eqref{S5-resp} shows two independent
parity and time reversal invariant expressions, build out of an odd number of
pseudovectors $\tilde A$. Therefore, anomalies of general fermionic particles
involve two parameters depending on their vector and axial
couplings\footnote{In Appendix \ref{app:coeffs} we collect anomaly formulas
  for general matter content.}. We fix
the first one to the value for the Dirac particle and let
the parameter $\a$ vary off the Dirac value $\a=1/3$ for later convenience.

As explained in Section \ref{sec:2.2}, the variation of response actions
gives the bulk currents and then the 3+1d covariant anomalies by inflow.
One finds
\begin{align}
  \!\!\!\!\!\!\!
  j^5=\frac{\d S}{\d A_5}=-2 \eps^{5\mu\nu\r\s}\de_\mu A_\nu \de_r \tilde A_\s
  \qquad \to\qquad
  \de_\mu J^\mu_{cov}& = -2 [dA d \tilde A]\,,
 \label{jcov-4d}
\\
  \tilde j^5=\frac{\d S}{\d\tilde A_5} \qquad\qquad\qquad\qquad\qquad
  \qquad \to\qquad
  \de_\mu \tilde J^\mu_{cov}& = - [dA d A] - 3\a [d \tilde A d \tilde A]\,,
 \label{jtcov-4d}
\end{align}
within our normalization $(e/2\pi)^2\to 1$. 

The gauge transformation $(A\to A +d\l, \tilde A \to \tilde A +d\tilde\l)$
of the bulk action \eqref{S5-resp} determines the WZW action
\be
S_{WZW}=-\int_{{\cal M}_4}\tilde\l ( d A d  A +\a d \tilde A d\tilde A)\,,
\label{S-WZW4d}
\ee
and its variation provides the anomalies of consistent currents,
\begin{align}
  \de_\mu J^\mu_{cons}& = 0\,,
  \label{jcons-4d}
  \\
  \de_\mu \tilde J^\mu_{cons}& = - [dA d A] - \a [d \tilde A d \tilde A]\,.
  \label{jtcons-4d}
\end{align}
Note that the order of $(A,\tilde A) $ factors in the first term
of the action \eqref{S5-resp} can be changed by adding 3+1d boundary terms;
the present form is vector gauge invariant, leading to a conserved consistent
vector current in agreement with
the hydrodynamic approach.

Finally, following Section \ref{sec:2.2}, the relation
between covariant and consistent currents is obtained by close inspection of
boundary terms in the variation of the bulk action 
as done in the 1+1d case \cite{arouca2022quantum}:
\begin{align}
  J^\mu_{cov}& = J^\mu_{cons} -2 [\tilde A d A]^\mu\,,
\\
  \tilde J^\mu_{cov}& = \tilde J^\mu_{cons} - 2\a [\tilde A d \tilde A]^\mu\,.
  \label{j-rel4d}
\end{align}
These expressions match the anomalies just found.

%-4.2-%%%%%%%%%%%%%%%%%%%%%%%%%%%%%%%%%%%%%%%%%%%%
\subsection{Hydrodynamics with axial current}
\label{sec:42}

The hydrodynamic description of 3+1d anomalies developed in
Ref.\cite{AbanovII} follows the same steps of the earlier 1+1d case.
The axial current is identified as the 3+1d helicity current
\eqref{top-curr}, which is coupled to its background as follows\footnote{
Note that the coupling to the topological current is expressed by a top form.}
\begin{align}
    S[\pi] = \int d^4x\, P(\pi-A) +\int_{{\cal M}_4} \tilde A(\pi-A)d(\pi+A) \,.
 \label{S-hydro4d}
\end{align}
The helicity current \eqref{top-curr} is rewritten by replacing $p=\pi-A$
and an additional term is introduced which was physically motivated in
\cite{AbanovII} and will be confirmed by the following analysis.

The admissible (diffeomorphism) variations of this action involves the
particle current,
\be
    \mathcal{J}^\mu = -\frac{\delta S}{\delta \pi_\mu} 
    = -\frac{\p P}{\p \pi_\mu} +
    \left[2\tilde A d\pi -(\pi-A)d\tilde A\right]^\mu\,,
\ee
which obeys the Carter-Lichnerowicz equation \eqref{dpi-100}.
Following the discussion of Section \ref{sec:311}, in 3+1 dimensions this
reduces to
the constraint $d\pi d\pi=0$; thus, the equations of motion are summarized by
\be
    \de_\mu {\cal J}^\mu=0, \qquad\qquad d\pi d\pi=0\,.
 \label{CL-4d}
\ee

A standard approach in hydrodynamics is that of introducing the
so-called Clebsch parameterization\footnote{See Appendix
  \ref{sec:litvar} for details.} for $\pi$ in terms of three scalar
functions $(\th,\a,\b)$, as follows:
\be
\pi=d\th+\a d\b, \qquad\qquad
\pi d\pi=d\th d\a d\b \,,
 \label{Clebsh}
\ee
which provide a local solution to the constraint $d\pi d\pi=0$.
However, contrary to 1+1d, this non-linear map complicates expressions
and cannot be extended globally because the parameters $(\th,\a,\b)$
develops singularities \cite{jackiw-rev}.  We shall, therefore,
continue using the momentum variable $\pi$, keeping in mind the
condition $d\pi d\pi=0$.

The consistent currents are readily evaluated
\begin{align}
  &J^\mu_{cons} = \frac{\delta S}{\delta A_\mu} =
    -\frac{\p P}{\p \pi_\mu} +[(\pi-A)d\tilde A +2\tilde A dA]^\mu =
    \mathcal{J}^\mu +\left[d\Big(2\tilde A(\pi-A)\Big)\right]^\mu\,,
\label{jcons-hy} \\
  &\tilde J^\mu_{cons} = \frac{\delta S}{\delta \tilde A_\mu} =
    [(\pi-A)d(\pi+A)]^\mu\,.
\label{jtcons-hy}    
\end{align}
Using equations of motion, we find
\begin{align}
    \de_\mu J^\mu_{cons} &= 0\,,
 \\
    \de_\mu\tilde J^\mu_{cons} &= -[dAdA]\,.
 \label{cons-div-100}
\end{align}

These results were obtained in the analysis of Ref. \cite{AbanovII};
however, comparison with the general expressions \eqref{jcons-4d},
\eqref{jtcons-4d} in the previous Section, shows that a term is
missing in the axial current, proportional to $[d\tilde A d\tilde A]$.
Namely, the hydrodynamic theory (\ref{S-hydro4d}) is able to reproduce
the anomalies correctly only for the parameter $\a=0$. In the
following Sections, we shall see how to obtain this missing piece, as
well as the contribution coming from curved metric backgrounds.

%-4.2.1-%%%%%%%%%%%%%%%%%%%%%%%%%%%%%%%%%%%%%%%%%%%%
\subsubsection{Free variations with generalized constraint}
\label{sec:421}

Paralleling the 1+1d analysis in Section \ref{sec:322}, we now
consider the standard variational problem for the action
\eqref{S-hydro4d}, by adding the constraint via a Lagrange multiplier.
The action is now
\begin{align}
  S[\pi] = \int_{{\cal M}_4} P(\pi-A) +\tilde A(\pi-A)d(\pi+A) +
  \psi (d\pi d\pi +\a d \tilde A d\tilde A)\,,
 \label{S-hy-psi}
\end{align}
where $\psi$ is pseudoscalar.

In this action, we also considered an extension of the constraint that
is motivated by the form of the WZW action \eqref{S-WZW4d}, with
parameter $\a$: this modification will be fully justified in the next
Section by matching the anomaly inflow from the 4+1d topological
theory.

The equations of motion obtained by varying freely over $\pi$ and
$\psi$ are:
\begin{align}
 \pi:\qquad & -\frac{\p P}{\p \pi_\mu} +
    \left[2(\tilde A-d\psi) d\pi -(\pi-A)d\tilde A\right]^\mu=0\,,
  \\
  \psi:\qquad &  \ d\pi d\pi+\a d \tilde A d\tilde A =0\,.
\label{eqm-4d-con}
\end{align}
Comparing with those obtained by
admissible variations \eqref{CL-4d} (for $\a=0$), one sees that
free variations again imply a `first integral' of the particle current
conservation:
\begin{align}
  & {\rm admissible:}\qquad\qquad \de_\mu {\cal J}^\mu=
\de_\mu\left(-\frac{\p P}{\p \pi_\mu} +
    \left[2\tilde A d\pi -(\pi-A)d\tilde A\right]^\mu\right) =0\,,
  \\
  & {\rm free:}\qquad\qquad\qquad\qquad
 {\cal J}^\mu \underset{eq. m.}{=}2[d\psi d\pi]^\mu\,.
\label{J-bad}
\end{align}

The parameterization \eqref{J-bad} of the current in terms of $\psi$
is general enough when $d\pi\neq 0$, i.e. in presence of fluid
vorticity. As fully described in Appendix \ref{sec:litvar}, it also
holds at points where $d\pi\to 0$ by considering a limiting procedure
where $d\psi\to\infty$, such that the current stays finite.  In
conclusion, free variation of the action \eqref{S-hy-psi} is
equivalent to admissible variation for $\a=0$, and extends to the case
$\a\neq 0$.

Let us show that the correct anomalies and conservation laws are
obtained. The vector current $J_{cons}$ has the same form as the one
of the previous Section \eqref{jcons-hy} and is conserved. The
consistent current $\tilde J^\mu_{cons}$ acquires a contribution from
the term parameterized by $\a$,
\be
    \tilde J^\mu_{cons} = [(\pi-A)d(\pi+A) +2\a d\psi d\tilde A]^\mu\,.
 \label{jtcons-al}
\ee
Its conservation law follows from \eqref{eqm-4d-con}
\be
\de_\mu\tilde J^\mu_{cons} = -[dAdA] -\a [d\tilde A d\tilde A]\,.
\label{jt-anom-corr}
\ee

The equation satisfied by the energy-momentum tensor is obtained by
considering a diffeomorphism transformation of the action. Since the
top form $\psi( d\pi d\pi +\a d \tilde A d\tilde A)$ is
diffeomorphism-invariant, it does not contribute to the
energy-momentum tensor, and the result is the same as
\eqref{deltaS-310}:
\be
 \de_\nu T^\nu_\mu = F_{\mu\nu}J^\nu_{cons} -
             A_\mu\p_\nu J^\nu_{cons} +\tilde F_{\mu\nu}\tilde J^\nu_{cons} -
             \tilde A_\mu\p_\nu \tilde J^\nu_{cons}\,.
\label{T-diff}
\ee
Using (\ref{jt-anom-corr}) and the 3+1d algebraic identity\footnote{
  See \eqref{4d-id} in Appendix \ref{sec:litvar}.}
$\tilde A_\mu [dAdA]=-2F_{\mu\nu}[\tilde AdA]^\nu$
we find the following expressions
\begin{align}
  \p_\mu T^\mu_\nu &= F_{\nu\mu}J^\mu_{cov}
                     +\tilde F_{\nu\mu}\tilde J^\mu_{cov}\,,
\label{T-102}
 \\
    J^\mu_{cov} &= J^\mu_{cons} -2[\tilde AdA]^\mu\,,
 \\
    \tilde J^\mu_{cov} &= \tilde J^\mu _{cons} -2\a [\tilde A d\tilde A]^\mu\,,
\end{align}
involving the covariant currents (cf.~\eqref{j-rel4d}).
Their anomalies have the correct form \eqref{jcov-4d}, \eqref{jtcov-4d}
which we reproduce here for convenience:
\begin{align}
    \de_\mu J_{cov}^\mu &=  -2[d\tilde A dA]\,, 
 \\
  \de_\mu \tilde J_{cov}^\mu &= -[dAdA] -3\a [d\tilde A d\tilde A]\,.
\label{4d-anom}
\end{align}

Therefore, we have shown that the hydrodynamic action
\eqref{S-hy-psi} provides a bosonic effective
description of 3+1d anomalies.
The field variables are the fluid momentum
$\pi$ and the new pseudoscalar $\psi$, obeying the gauge conditions
\be
p=\pi -A, \qquad\qquad \tilde q=d\psi -\tilde A, \qquad\qquad
{\rm gauge\ invariant} .
\label{g-cond2}
\ee

%-4.3-%%%%%%%%%%%%%%%%%%%%%%%%%%%%%%%%%%%%%%%%%%%%

%---------------------------------------------------
\subsection{4+1 Dimensional `hydrodynamic' topological theories}
\label{sec:43}

The derivation of the 2+1d bulk theory in Section \ref{sec:23} started from
the introduction of `hydrodynamic' gauge fields that are dual to
conserved currents, such as $j=*d a$. In 4+1 dimensions, the corresponding
expressions, $j=*d c$, involve 3-form fields $c$, which could be independent
degrees of freedom or polynomial of lower-order forms.
The anomaly inflow relations \eqref{jcov-2d}, \eqref{jtcov-2d}
then imply 3+1d currents parameterized as $J_{cov}=*c$.

The hydrodynamic currents obtained in previous Sections,
\eqref{jcons-hy}, \eqref{jtcons-al}, once evaluated on the 3+1d
equations of motion, have the form
\begin{align}
  & * J_{cov} \underset{eq.m.}{=}
    2 (d\psi-\tilde A)d\pi+ 2(\pi-A)d\tilde A,
\label{jcov4d-eqm}  \\
  & * \tilde J_{cov}\underset{eq.m.}{=} (\pi-A)d(\pi+A) +
    2\a (d\psi -\tilde A)d \tilde A \,,
\label{jtcov4d-eqm}
\end{align}
where the fluid momentum $\pi$ is constrained by
$d\pi d\pi +\a d\tilde A d\tilde A=0$.
These expressions\footnote{Note that they are explicitly
  gauge invariant, being functions of the
  quantities $(\pi-A)$ and $(d\psi-\tilde A)$, see \eqref{g-cond2}.}
are polynomials of the field $p=\pi -A$ and its axial counterpart
$\tilde q=d\psi -\tilde A$.

Therefore, we take a conservative approach and search for 4+1d
topological actions which reproduce the hydrodynamic currents and thus
are polynomials in one-form fields; we also expect cubic expressions
implying quadratic equations of motion \eqref{eqm-4d-con}. We shall
find one such theory that can be considered as the minimal one,
together with other instances describing more general dynamics and
additional degrees of freedom.

The derivation starts by reconsidering the 2+1d theory, in the chiral case
for simplicity. We rewrite the Chern-Simons action \eqref{CS},
up to a boundary term (c.f. \ref{2+1d-act-pi})), as
\be
S=\int_{{\cal M}_3} \frac{1}{2} ada +adA=
\frac{1}{2} \int_{{\cal M}_3} \pi d\pi -  AdA \,,
\qquad \qquad (\pi=a+A).
\label{CS-pi}
\ee
It is apparent that the stationary point is quadratic in $\pi$ and thus
the equation of motion is linear, $d\pi=0$.
The generalization to 4+1 dimensions is the cubic expression
\be
S=\frac{1}{6}\int_{{\cal M}_5} \pi d\pi d\pi- AdAdA, \qquad\qquad (\pi=a+A),
\label{S-weyl2}
\ee
leading to the desired quadratic condition $d\pi d\pi=0$.

Note that variations of these actions can be equivalently taken with respect
to $a$ and $\pi$ since $A$ is inert. Furthermore, the current
can be defined by varying the action at fixed $a$ or $\pi$: the two
resulting expressions become equal once evaluated on the
equations of motion (as they should).
In the following, it is more convenient to focus on the dynamics of
$\pi$ in both bulk and boundary actions.

Note also that the actions \eqref{CS-pi}, \eqref{S-weyl2} are
differences of two Chern-Simons theories for gauge fields $\pi$ and
$A$ which transform in the same way,\footnote{This gauge condition has
  been found before, see \eqref{g-cond2}.}
$A\to A +d\l, \pi\to\pi+d\l$. These expressions are called
`transgression forms' in the mathematical literature
\cite{Nakahara:Geometry}, where they are made gauge invariant by
adding a boundary term, in the same spirit of our approach.
Transgression forms have already been introduced in hydrodynamics
\cite{Jensen-transgr,Haehl:2013hoa}, within an approach using
different fluid variables as fundamental fields.

%-4.3.1-%%%%%%%%%%%%%%%%%%%%%%%%%%%%%%%%%
\subsubsection{Two-fluid theory}
\label{sec:431}

The topological action which realizes the precise
doubling of the chiral action \eqref{S-weyl2} is:
\begin{align}
  S_{II} =&\int_{{\cal M}_5} \tilde \pi d\pi d\pi- \tilde A dAdA
     +\a \tilde \pi d\tilde\pi d\tilde \pi- \a\tilde A d\tilde A d\tilde A
 \nonumber   \\
 & +\int_{{\cal M}_5} \!\!\!\!\! P(\pi-A,\tilde\pi-\tilde A) +
   \tilde A \left[ (\pi-A)d(\pi+A)+
   \a (\tilde \pi -\tilde A)d(\tilde \pi +\tilde A)\right] \,,
       \label{S_II}
    \\
 &    \qquad\qquad \pi=p+A, \qquad \qquad \tilde\pi=\tilde q +\tilde A\,,
 \label{S-fluids}
\end{align}
involving two transgression forms and respecting parity and time
reversal invariances. In this action the symmetry between the two
fields $p,\tilde q$ and corresponding fluids is manifest. For this
reason, it will be called the `two-fluid' theory. The 3+1d boundary
action is obtained following the same strategy of Section
\ref{sec:23}: it involves the pressure dynamic term that can depend on
both fluid variables, and the axial coupling whose expression is
consistent with overall bulk-boundary gauge invariance (having
established that the combinations $\pi-A$ and $\tilde\pi-\tilde A$ are
gauge invariant).

The bulk equations of motion following from (\ref{S_II}) are 
\be
d\pi d\tilde\pi=0\,, \qquad \qquad d\pi d\pi +
3\a d\tilde\pi d\tilde\pi=0\,.
 \label{eqm-2fl}
\ee
The 3+1d covariant currents determined by anomaly inflow
are 
\begin{align}
  & *J_{cov}= 2 \tilde \pi d\pi - 2 \tilde A dA + d \tilde b\,,
\label{j-2fluid}    \\
  &  *\tilde J_{cov}=(\pi-A)d(\pi+A) +
    3\a (\tilde\pi-\tilde A)d(\tilde\pi+\tilde A)  +d b\,,
\label{jt-2fluid}
\end{align}
where $b,\tilde b$ are undetermined two-form fields.  The conservation
of these currents evaluated on the bulk equations of motion lead to
the expected anomalies of Section \ref{sec:red4+1top}, now realized by
the two fluids on equal footing. It can also be shown that the
boundary equations of motion lead to the correct conservation
equations for consistent currents and stress tensor.

We note that the expressions of currents \eqref{j-2fluid}, \eqref{jt-2fluid}
differ from those found earlier in hydrodynamics \eqref{jcov4d-eqm},
\eqref{jtcov4d-eqm}, in particular for the
presence of two independent fluid momenta \eqref{S-fluids}.

Actually, the topological action \eqref{S_II} describes systems where
chirality is conserved, which are physically different from those
considered so far. In this work, we have been assuming a single Dirac
fluid described by $\pi$, where the other variable $\tilde\pi= d\psi$
does not have a proper dynamics, i.e. it is `dragged', and can
describe an irrotational fluid component, at most. We imagine that
fermion interactions lead to a dynamical effective mass and
non-conservation of fermion chirality. One check of this fact is given
in Appendix \ref{app:C}, where we discuss the behavior of Dirac fluids
in a chiral background, e.g. $A=\tilde A$: we find that the two chiral
currents \eqref{chi-split}), $J_\pm=J \pm \tilde J$,
do not decouple from each other. The properties of two-fluid hydrodynamics
\eqref{S_II} as well as its chiral half will be analyzed in future
investigations.

%-4.3.2-%%%%%%%%%%%%%%%%%%%%%%%%%%%%%%%%%
\subsubsection{Single-fluid theory and Dirac hydrodynamics}
\label{sec:432}

We now consider the reduction of the two-fluid theory \eqref{S_II}
assuming that the axial momentum is irrotational and that the pressure is
only function of vector momentum $\pi-A$,
\be
    \tilde\pi=d\psi, \qquad\qquad P=P(\pi-A), 
\ee
which will be referred to as the `single-fluid' theory.

In this case, part of the bulk action in \eqref{S_II} is a total
derivative and reduces to a boundary term so that one is left with
the expression
\begin{align}
    S_I=  &- \int_ {{\cal M}_5} \tilde AdAdA+\a \tilde Ad\tilde A d\tilde A
 \nonumber\\
    &+\int_{{\cal M}_4} P(\pi-A) +\tilde A(\pi-A)d(\pi+A) +
    \psi(d\pi d\pi+\a d\tilde A d\tilde A)\,.
 \label{S-bb-4d}
\end{align}
We recognize that the 3+1d term is equal to the hydrodynamic action
\eqref{S-hy-psi} introduced in Section \ref{sec:421}, where $\psi$ is
the Lagrange multiplier enforcing the equation of motion
\eqref{eqm-4d-con}.  The 4+1d piece is the response action
\eqref{S5-resp} not participating to the boundary dynamics.

Therefore, we have obtained the 3+1d hydrodynamic theory of Section
\ref{sec:42} from the study of bulk topological theories and their
degrees of freedom, under the assumption of a single dynamic
fluid. This nonetheless includes the additional pseudoscalar variable
$\psi$ with respect to the 3+1d Euler hydrodynamics described in
Ref.~\cite{AbanovII}. This additional degree of freedom is crucial for
obtaining the most general anomaly characterized by $\alpha\neq 0$.

We remark that in this theory the anomaly inflow relations only
  involve the 4+1d response action \eqref{S5-resp} and determine the
  anomalies of covariant currents \eqref{jcov-4d},\eqref{jtcov-4d}.
  The correspondence between bulk and boundary hydrodynamic fields
  is rather clear and can be partially checked through the form
  of currents.  

%-4.3.3-%%%%%%%%%%%%%%%%%%%%%%%%%%%%%%%%%
\subsubsection{Theory involving higher-form fields}
 
Earlier in this Section we pointed out that the description of matter currents
in 4+1d and 3+1d naturally suggests the inclusion of 2- and 3-form hydrodynamic
fields $b,c$ into the topological action. They were not considered
in the two previous theories, which only involve the one-form fields
expressing fluid momenta: in particular, the single-fluid case
correctly matches the expected hydrodynamic theory, in which
vorticity and helicity are polynomial of fluid momentum/velocity.

Nonetheless, we now present another 4+1d topological theory which is
consistent with Dirac anomalies and makes use of 3-form fields for
parameterizing currents. In this theory, fluid velocity and vorticity are
described by two independent degrees of freedom. Let us consider the following
hydrodynamic topological action
\begin{align}
  S_{IIb}=\int_{{\cal M}_5}\tilde c  d(p+A) +c  d(\tilde q +\tilde A) +
  \tilde q  d p  d p +\alpha  \tilde q d\tilde q   d\tilde q\,,
 \label{action-90}
\end{align}
where the $A, \tilde A$ backgrounds couple to dual 3-form currents
$c,\tilde c$, while the $p,\tilde q$  1-forms realize 4+1d Chern-Simons terms. 
This is the natural generalization of the 2+1d theory
\eqref{CS}, and is similarly characterized by quadratic stationary points.

The solution of the bulk equations of motion are
\begin{align}
  c:\quad  &\ \tilde q=-\tilde A +d\psi, \qquad i.e.\qquad \tilde\pi=d\psi\,,
 \\
 \tilde c:\quad  &\ p=-A+d\th, \qquad i.e.\qquad \pi=d\th\,, 
 \\
  p: \quad &\ \tilde c= -2 pd \tilde q+ d b= 2(d\th-A)d\tilde A +db,
    \\
  \tilde q:\quad &\ c=-pdp-3\a \tilde qd\tilde q +d b
  = (d\th -A)dA +3\a (d\psi -\tilde A)d\tilde A +d \tilde b\,,
\end{align}
where $b,\tilde b$ are undetermined 2-form fields.
The hydrodynamic action \eqref{action-90} evaluated on the solution
of equations of motion gives the response action \eqref{S5-resp}, as needed.

The anomaly inflow relations determine the following
functional form of covariant 3+1d currents
\begin{align}
  &  * J_{cov} = \tilde c \underset{eq.m.}{=}
    = 2(d\th -A)d\tilde A + d  b\,,
\label{j-c} \\
  &  *\tilde J_{cov}=c \underset{eq.m.}{=}
    (d\th -A)dA +3\a (d\psi -\tilde A)d\tilde A +d \tilde b\,.
\label{jt-c}
\end{align}
Gauge invariance requires
\be
c,\qquad \tilde c, \qquad p= d\th -A, \qquad \tilde q= d\psi -\tilde A,
\qquad {\rm gauge\ invariant\ on\ } {\cal M}_4
\ee
which generalize earlier gauge conditions. These establish that
 $\th,\psi$ are compensating scalar/pseudoscalar gauge field.

 The expressions of currents \eqref{j-c},\eqref{jt-c} have
 rather interesting physical consequences:
 \begin{itemize}
 \item
   They identify a 3+1d hydrodynamics characterized by a pair of
   scalar fields and a pair of two-form fields $b, \tilde b$. 
The latter parameterize the 3+1d currents in absence of backgrounds.
 \item
   The currents \eqref{j-c},\eqref{jt-c}
   match the expressions found for the two-fluid theory,
   \eqref{j-2fluid}, \eqref{jt-2fluid}, evaluated for
   $d\pi=d\tilde\pi=0$. In that case, the contribution by the two-form
   fields was not actually necessary, but here is rather important.
\item 
 The Dirac anomalies are reproduced by the expressions
 \eqref{j-c}, \eqref{jt-c} irrespectively of the value of
 $b,\tilde b$ fields. The hydrodynamics with $b=\tilde b=0$ describes
 a system with two irrotational fluids, owing to $d\pi=d\tilde\pi=0$,
 which is static in absence of backgrounds. This minimal theory
 shows that the requirement of reproducing 3+1d anomalies is not
 very strong. The response is entirely anomalous and
 quadratic in the background fields: e.g. in a static magnetic field
 causing the dimensional reduction, it reproduces the physics of 1+1d systems
\cite{arouca2022quantum}.
\end{itemize}

We conclude that the topological theory \eqref{action-90} identifies
an interesting non-standard hydrodynamics. The 3+1d action is obtained
as in earlier cases by adding a dynamic pressure term,
$P=P(d\th -A,d\psi -\tilde A, db, \tilde db) $,
in general depending on all fields, and the coupling to
vector and axial backgrounds for reproducing the consistent currents.
Again anomalies are independent of the form of the pressure.
The minimal fluid solutions with $b=\tilde b=0$ do not involve any vorticity, 
while the general one include vorticities which are parameterized by
independent degrees of freedom.
The analysis of this theory as well as that in Section \ref{sec:431}
requires an independent investigation.

We remark that the list of 4+1d topological theories presented here and 
corresponding generalized hydrodynamics is not meant to be exhaustive.
Among further possibilities, it would be interesting to study
the hydrodynamics of chiral-only (Weyl) fermions, which can be
obtained from the two-fluid theory by identifying the two halves.

Interacting phases of matter in 3+1 and 4+1d can give rise to
monodromy phases between point-particles and extended excitations that
are also described by topological theories involving higher forms
\cite{Cho:2010rk} \cite{Andreucci:2019ltx}.
For example, the addition of a BF term $\int \tilde q d b$ in the
3+1d action can describe anyonic monodromies of vortex lines and particles.
Finally, associated to extended excitations there are higher-form
extended symmetries that allow for further possible terms in
the topological actions \cite{Gaiotto-higher}.

The complete description of higher-form excitations requires the introduction
of corresponding higher form backgrounds $B,C$, leading to
 higher anomalies. The response of systems with respect to
these new external fields may help in distinguishing among
the various topological theories compatible with Dirac
anomalies in  4+1 dimensions.

These generalizations of the hydrodynamic/effective bosonic theory
will be considered in future works.  The correspondence between
hydrodynamic, bosonization of fermions and topological theories
in one extra dimension is the main result of this work. It is apparent
that insight from topological theories in 4+1d can lead to very
interesting generalizations of hydrodynamics and corresponding bosonic
effective descriptions of anomalies.

%-5-%%%%%%%%%%%%%%%%%%%%%%%%%%%%%%%%%%%%
\section{3+1 Dimensional axial-gravitational anomaly }
\label{sec:5}

%-5.1-%%%%%%%%%%%%%%%%%%%%%%%%%%%%%%%%%%%%

\subsection{Axial anomaly on Riemannian geometries}
\label{sec:51}

The 3+1d hydrodynamic theory is now considered in the presence of a
gravitational background, expressed by vierbein and spin connection
$(e^a_\mu, \w_{\mu}^{ab}$). We shall first consider Riemannian
geometries (vanishing torsion) and later discuss the general case.

Fermions do not have a purely gravitational anomaly
($\de_\mu T^{\mu\nu}\neq 0$ only occurs in dimensions
$d=2,6, 10,\dots$), but show a gravitational contribution to the axial
anomaly \cite{LAG-Witten}
\be
    D_\mu \tilde J^\mu_{cov}=-\frac{\b}{4} \epsilon^{\mu\nu\r\s}
    R_{\mu\nu}{}^a_{\ b}\; R_{\r\s}{}^b_{\ a} = -\b [\Tr( R^2)]\,.
 \label{grav-ax}
\ee
This expression contains the Riemann 2-form
$R^a_{\ b}=\frac{1}{2} R_{\mu\nu}{}^a_{\ b}dx^\mu \wedge dx^\nu$,
expressed as
$R^a_{\ b}=d\w^a_{\ b} +\w^a_{\ c}\w^c_{\ b} $
in terms of the spin-connection 1-form, 
$\w^\a_{\ b}=\w^a_{\mu b}dx^\mu$.
The trace  is over the local Lorentz  indices $a,b$, and
$D_\mu$ is the covariant derivative wrt the gravitational
background.\footnote{ Note the covariant form of the antisymmetric tensor
  $\epsilon^{\mu\nu\cdots\l} = \eps^{\mu\nu\cdots\l}/|e|$,
  where the metric determinant is $\sqrt{- g}=|e|$. }
The integral of the mixed axial-gravitational anomaly \eqref{grav-ax}
is proportional to the pseudoscalar
4d Pontryagin topological invariant.\footnote{There is no analog in 1+1d since
  the curvature is a scalar quantity.} The anomaly coefficient 
for a Dirac fermion is $\b= 1/96\pi^2 \to 1/24$, in our conventions.

The anomaly \eqref{grav-ax} can be described by the following 4+1d
addition to the hydrodynamic theory \eqref{S-bb-4d}
\be
\D S_I = \b\int_{{\cal M}_5} (d\psi-\tilde A) \Tr (R^2)\,.
\label{grav-topo}
\ee
Since the curvature 4-form obeys
$\Tr (R^2)=d \W_{CS,3}(\w)=d\Tr (\w d\w +2\w^3/3)$, the mixed
Chern-Simons 5-form $\W_{CS,5}(A,\w)=A \; \Tr (R^2)$ verifies
$d\W_5=F\Tr (R^2)$, which is the topological invariant 6-form
appearing in the 5+1d index theorem for mixed gauge-gravitational
anomalies. Thus, the expression \eqref{grav-topo} is consistent with
the literature on anomalies and their relations in various dimensions
\cite{treiman2014current,bertlmann2000anomalies}.
Earlier hydrodynamic approaches including gravitational anomalies
can be found in \cite{Nair:2011,Volovik-PRD}.

The 4+1d current obtained by varying the action \eqref{grav-topo}
w.r.t. $\tilde A$ reproduces the contribution \eqref{grav-ax} by anomaly
inflow, following the same steps an in Section \ref{sec:431}.
Furthermore, the WZW action associated to $\D S_I$ is,
\be
    \D S_{WZW}=-\b\int_{{\cal M}_4} \tilde \l \Tr (R^2)\,,
 \label{grav-WZW}
\ee
which determines a consistent anomaly equal to \eqref{grav-ax} --
the gravitational contributions to
$D_\mu\tilde J^\mu_{cov}$ and $D_\mu\tilde J^\mu_{cons}$ are equal.

The 3+1d hydrodynamic action \eqref{S-bb-4d} takes the following
form in presence of vector, axial and gravity backgrounds
\begin{align}
  S=&\int_{{\cal M}_4}d^4x\; |e| P(p) +
     \int_{{\cal M}_4} \tilde A(\pi-A)d(\pi+A) +
\psi\left(d\pi d\pi+\a d\tilde A d\tilde A +\b \Tr (R^2)\right)
\nonumber\\
    &- \int_ {{\cal M}_5} \tilde A\left(dAdA+\a d\tilde A d\tilde A+
      \b \Tr (R^2)\right)\,.
\label{S-hydrob}
\end{align}
One recognizes the additional $\Tr (R^2)$ terms and the metric
determinant multiplying the pressure term, while exterior derivatives
are unaffected by the Levi-Civita connection, owing to
$\G_{\a\b}^\l=\G_{\b\a}^\l$. Similarly, Lie derivatives can be
expressed in terms of both ordinary and covariant derivatives.

It follows that earlier expressions of $J, \tilde J$ and $T^\nu_\mu$ are
unchanged. We rewrite them for convenience
\begin{align}
  T^\nu_\mu &= -\frac{\p P}{\p \pi_\nu}(\pi_\mu-A_\mu)+P\delta^\nu_\mu\,,
 \label{T-55}\\    
    J^\nu_{cov} &= -\frac{\p P}{\p \pi_\nu} +\Big[(\pi-A)d\tilde A\Big]^\nu \,,
 \label{J-55}\\
  \tilde J^\nu_{cov} &= \Big[(\pi-A)d(\pi+A) -
                       2\alpha (\tilde A-d\psi)d\tilde A\Big]^\nu \,.
 \label{Jt-55} 
\end{align}

The $\pi$ equation of motion is also unchanged, while that of $\psi$
acquires the $R^2$ term
\begin{align}
 \pi:\qquad & -\frac{\p P}{\p \pi_\mu} +
    \left[2(\tilde A-d\psi) d\pi -(\pi-A)d\tilde A\right]^\mu=0\,,
  \\
\psi:\qquad &  \ d\pi d\pi+\a d \tilde A d\tilde A +\b \Tr (R^2)=0\,.
\end{align}

It follows that currents now obey the expected anomalous equations
\begin{align}
    D_\nu J^\nu_{cov} &= -\Big[2d\tilde A dA\Big]\,,
 \label{dJ-55}\\
  D_\nu \tilde J^\nu_{cov} &= -\Big[dA dA+3\alpha d\tilde A d\tilde A
                             +\beta\tr (R^2)\Big]\,,
 \label{dJt-55}
\end{align}
in presence of all three backgrounds.

%-5.2-%%%%%%%%%%%%%%%%%%%%%%%%%%%%%%%%%%%%
\subsection{The spin current}
\label{sec:52}

The stress tensor \eqref{T-55} has been obtained by diffeomorphism
invariance and thus is defined in terms of metric variations of the action
\be
    \d S=\frac{1}{2}\int |e| T_{\mu\nu}\d g^{\mu\nu}\,.
 \label{Te-def}
\ee

When using the variables $(e_\mu,\w_\mu)$ it is possible to consider independent
variations with respect to each background, as follows
\be
\d S=\int |e| \left( t_a^\mu \d e^a_\mu +
 \frac{1}{2}{\cal S}^\mu_{ab} \d \w^{ab}_\mu \right)\,,
\label{t-def}
\ee
which define a new response given by the spin current ${\cal S}_{ab}^\mu$
and another definition of the stress tensor $t^\mu_a$.
The backgrounds $(e,\w)$ and corresponding currents $(t,{\cal S})$ are
independent variables only for geometries with torsion.  In absence of
it, the spin connection can be expressed in terms of vierbeins,
$\w_\mu^{ab}=\w_\mu^{ab}(e)$, using the metric compatibility condition
$D_\l e^a_\mu=0$.  Therefore, for Riemannian geometries the stress
tensor $T$ in \eqref{Te-def} is well defined while $t, {\cal S}$ can
be redefined by shifting terms among them. As explained in
Ref.\cite{yarom-torsion}, these redefinitions can be seen as
improvements of the stress tensor $t$ by derivatives of ${\cal S}$.

In absence of backgrounds, both the stress tensor and the angular
momentum current $J^{\mu,\a\b}$ are conserved
\be
 J^{\mu,\a\b}=x^{[\a} T^{\mu\b]} 
    + {\cal S}^{\mu,a\b}\,,  \qquad\qquad
    \de_\mu J^{\mu,\a\b}=x^{[\a}\de_\mu T^{\mu\b]} + T^{[\a\b]}
    + \de_\mu {\cal S}^{\mu,a\b}=0\,.
 \label{J-cons}
 \ee
Therefore, the spin current is conserved
 when the stress tensors is both conserved and symmetric. For a
 Riemannian metric, the stress tensor \eqref{T-55}, defined by
 \eqref{Te-def}, is indeed symmetric; the other quantity obtained
 in \eqref{t-def}, $t^{\mu\nu}=t^\mu_a E^{a\nu}$, multiplied by the inverse
 vierbein, can be non-symmetric. Note, however, that the conservation
 laws of  $T$ and ${\cal S}$ are  modified
 in presence of external backgrounds and corresponding anomalies, as
 shown in the following analysis.

 Keeping in mind these facts, we now derive the spin current in
 presence of a Riemannian background. The case of geometries with
 torsion will be considered later.  We first vary the 4+1d topological
 action \eqref{grav-topo},
\be
    ({\cal S}_{(5)})^\mu_{ab} = \frac{2}{|e|}\frac{\d S_5}{\d \w_\mu^{ba}}=
    \frac{2\b}{|e|} \frac{\d}{\d \w_\mu^{ba}}\int_{{\cal M}_5}
    (d\psi- \tilde A)\Tr(R (d\w+\w^2))\,,
 \label{5d-spin-curr}
\ee
leading to the following terms
($\hat A=\tilde A -d\psi$),
\be
    d(\hat A R)_{ab} - \hat A (\w R)_{ab} + \hat A (R\w)_{ab}=
    d\hat A\; R_{ab} - \hat A( d R +\w R-  R\w)_{ab}=
    d \hat A\; R_{ab} \,,
\ee
as due to the Bianchi identity\footnote{This Bianchi identity also holds
  in presence of torsion.} $DR=dR +\w R-R\w=0$.

The anomaly inflow determines the following anomaly for
the 3+1d covariant spin current
\be
    ({\cal S}_{(5)})^5_{ab}=\left(D_\mu {\cal S}^\mu_{cov}\right)_{ab}=
    - 2\b \epsilon^{5\mu\nu\r\s}\de_\mu\hat A_\nu R_{\r\s, ab}=
    - 2\b \epsilon^{5\mu\nu\r\s}D_\mu(\hat A_\nu R_{\r\s, ab})\,,
 \label{s-anom}
\ee
where the last identity follows again from the Bianchi identity.
The current itself is\footnote{Note that this current is covariant
  wrt Lorentz indices, and invariant for
  the Abelian part due to the presence of $\psi$.
Components of dual forms are evaluated with the covariant epsilon tensor.}
\be
        {\cal S}^\mu_{ab, cov}=4 \b \left[ (d\psi-\tilde A) R_{ab} \right]^\mu
    +\cdots ,
 \label{S-cov}
\ee
where the dots are  `classic' terms that are not determined by the
bulk topological theory. 
Actually, contrary to the currents introduced earlier, the spin
current on the boundary might be non-conserved classically (even in
the absence of anomalies). Thus, for the time being we should consider
the expression \eqref{S-cov} as only the anomalous part of the spin current.

Next, we vary the 3+1d hydro action \eqref{S-hydrob} and observe that
the spin connection only appears in the anomalous $\Tr (R^2) $
term. Actually, terms involving differential forms are independent of
the metric. Repeating the same steps of the variation of
\eqref{5d-spin-curr}, we find
\be
    {\cal S}^\mu_{ab,cons}=4\b [d\psi \; R_{ab}]^\mu\,,
 \label{S-cons}
\ee
which obeys $D_\mu{\cal S}^\mu_{cons}=0$ due to the Bianchi identity.
This result can also be obtained by varying the WZW action \eqref{grav-WZW}
under a local Lorentz transformation $\d\L_{ab}$,
\be
    \d_\L S_{WZW}=\int_{{\cal M}_4} \d \L_{ab}
    \left( D_\mu {\cal S}^\mu\right)_{ab, cons}= -\b
    \d_\L \int_{{\cal M}_4} \tilde\l \Tr R^2 =0\,.
 \label{gWZW-var}
\ee

We observe that:

i) The consistent spin current is (covariantly) conserved, i.e.
non-anomalous;

ii) There are no additional (classic) terms in \eqref{S-cov}. Namely, the
spin current is entirely due to anomaly for Riemannian geometries;

iii) The result for the consistent spin current \eqref{S-cons} remains
valid for vanishing backgrounds $A=\tilde A =0$.

The relation between the two currents, consistent and covariant, is clearly
\be
    {\cal S}^\mu_{ab, cov}={\cal S}^\mu_{ab, cons}-
    4 \b \left[ \tilde A R_{ab} \right]^\mu\,,
 \label{S-diff}
\ee
and the anomalous conservation law is given by \eqref{s-anom}.

Let us now discuss the stress-tensor conservation due to the presence
of the curvature term in the 3+1d part of the action
\eqref{S-hydrob}. From diffeomorphism invariance, one finds the
expression
\begin{align}
    D_\nu T^\nu_\mu= & F_{\mu\nu} J^\nu_{cons}  +
    \tilde F_{\mu\nu} \tilde J^\nu_{cons}  -A_\mu D_\nu J^\nu _{cons}
                       - \tilde A_\mu D_\nu \tilde J^\nu_{cons}
   \nonumber\\
 & + \frac{1}{2}  \Tr \left(D_{[\mu} \w_{\nu]}{\cal S}^\nu_{cons} -
                       \w_\mu D_\nu{\cal S}^\nu_{cons}\right)\,,
 \label{T-110}
\end{align}
which differs from \eqref{T-diff} of Section \ref{sec:42} for the last
term and the covariant form of the Lie derivative $\d_\eps$. In the
case $\b=0$, the equation \eqref{T-110} can be rewritten as
\eqref{T-102},
\be
    D_\nu T^\nu_\mu= F_{\mu\nu} J^\nu_{cov}  +
    \tilde F_{\mu\nu} \tilde J^\nu_{cov}\,.
 \label{T-112}
\ee

These terms are also present for $\b\neq 0$, the expressions of
$T, J$ being unchanged.  There are two further pieces:

i) the contribution to
$-\tilde A_\mu\de_\nu \tilde J^\nu =\b \tilde A_\mu [\Tr R^2]$ due to the
gravitational anomaly;

ii) the variation of the action wrt the $\w_\mu$ background,
expressed in terms of the consistent current ${\cal S}_{cons}^\nu$.

These additions read,
\begin{align}
    \D_\mu&=   \b \tilde A_\mu [\Tr R^2]+  \frac{1}{2}              
            \Tr \left(D_{[\mu} \w_{\nu]}{\cal S}^\nu_{cons} -
            \w_\mu D_\nu{\cal S}^\nu_{cons}\right)
 \nonumber\\                                                                    
    & = \b \tilde A_\mu [\Tr R^2]+ 2\b \Tr(R_{\mu\nu}[ d\psi\; R]^\nu)
 \nonumber \\
    & = 2\b \Tr ( R_{\mu\nu}[(d\psi -\tilde A)R]^\nu)
    = \frac{1}{2}\Tr(R_{\mu\nu}{\cal S}_{cov}^\nu)\,.
 \label{cons-rev}
\end{align}
In the second line, we inserted the expression \eqref{S-cons} for
${\cal S}_{cons}^\nu$, that is conserved; in the third line we 
used the 3+1d identity (cf. \eqref{4d-id}),
$\tilde A_\mu \Tr [RR] =- 2 \Tr (R_{\mu\nu}[\tilde A R]^\nu)$.

In conclusion, the stress tensor equation in presence of the gravitational
background reads
\be
    D_\nu T^\nu_\mu= F_{\mu\nu} J^\nu_{cov}  +
    \tilde F_{\mu\nu} \tilde J^\nu_{cov} +
    \frac{1}{2}\Tr(R_{\mu\nu}{\cal S}^\nu_{cov})\,,
 \label{T-113}
\ee
where the expressions of stress tensor and covariant currents are
given in \eqref{T-55},\eqref{J-55} and \eqref{Jt-55}.

The contribution from the curved background nicely matches the form 
of other gauge backgrounds, corresponding to (generalized) Lorentz
forces. As it was explained above, the spin current (\ref{S-cov})
does not have classic contributions and can be written as
\be
    {\cal S}^\mu_{ab, cov}=4 \b \left[ (d\psi-\tilde A) R_{ab} \right]^\mu .
 \label{S-cov2}
\ee
Note that there is a difference in the fact that electric currents
have classical terms (see (\ref{J-55},\ref{Jt-55})), while the spin
current is only induced by quantum effects, leading to a force (last
term of \eqref{T-113}) proportional to the square of the curvature.

%-5.3-%%%%%%%%%%%%%%%%%%%%%%%%%%%%%%%%%%%%%%%
\subsection{Extension to geometries with torsion}
\label{sec:53}

% -5.3.1-%%%%%%%%%%%%%%%%%%%%%%%%%%%%%%%%%%%%%%%

\subsubsection{Relation between axial and spin currents 
for free Dirac fermions}
\label{sec:531}

We start by recalling some properties of free Dirac theory and setting
the notation. The stress tensor $T^{\mu\nu}$ and the total angular
momentum current
$J^{\mu,\a\b}$ are given by bilinears of the spinor field $\Psi$
\be
    T^{\mu\nu}=\frac{i}{2}\bar\Psi \g^\mu \overleftrightarrow{\de^\nu} \Psi,
    \qquad\qquad J^{\mu,\a\b}=x^{[\a} T^{\mu\b]} +
    \bar\Psi\frac{1}{4}\left\{\g^\mu,\s^{\a\b}\right\}\Psi
    =L^{\mu,\a\b}+{\cal S}^{\mu,\a\b}\,,
 \label{total-J}
\ee
where $L$ and ${\cal S}$ are the orbital and spin currents, respectively.
\footnote{Antisymmetrization of indices is indicated by square brackets,
  e.g. $F_{\mu\nu}=\de_{[\mu}A_{\nu]}$.}
The matrix $\s^{\a\b}$ is the generator of Lorentz rotations on spinors and
gamma matrices $(\g^\mu)_{ij}$,
\begin{align}
&  S(\W)\g^\mu S^{-1}(\W)=(\W^{-1})^\mu_{\ \nu}\g^\nu,
  \\
&\W_{\mu\nu}\sim \eta_{\mu\nu}+\L_{\mu\nu}\,, \qquad
  S(\W)_{ij}\sim I-\L_{ab}\frac{i}{4}(\s^{ab})_{ij}\,,
\qquad  (\s^{ab})_{ij}=\frac{i}{2}[\g^a,\g^b]_{ij}\,.
\end{align}

Our conventions for the gamma matrices can be summarized as follows
\be
  \{\g^\mu,\g^\nu\}=2\eta^{\mu\nu}=2\; {\rm diag}(-1,1,1,1),
  \qquad\quad \g^5=\g_5=i\g^0 \g^1\g^2\g^3 \,.
  \ee
They can be used to prove the following identity
\be
\frac{i}{8}\left\{\g_\a,[\g_\b,\g_\s]\right\}=\frac{1}{2}
\eps_{\a\b\s\mu}  \g^5\g^\mu\,,
\qquad\qquad (\eps^{0123}=-\eps^{1230}=1)\,.
\label{dirac-id}
\ee
Actually, the expression with three gammas in the lhs
vanishes for any pair of equal indices, in
some cases explicitly, in other cases due to $-(\g^0)^2=(\g^i)^2=1$.
It is also antisymmetric for any exchange of indices, thus mapping into
the expression involving $\g^5$.

The coupling of the spin connection to the spin current is as follows
\be
\w_\mu^{ab}\;{\cal S}^\mu_{ab}=
\w_\mu^{ab}\;\bar\Psi\frac{i}{8}\left\{\g^\mu,[\g_a,\g_b]\right\}\Psi\,.
\label{dirac-spin}
\ee

In the case of a flat metric $e^a_\mu=\d^a_\mu$,
we can freely pass from Latin to Greek indices and
use the gamma-matrix identity \eqref{dirac-id} to find the relations
\be
{\cal S}^{\mu,ab}=\frac{1}{2}\eps^{\mu a b \s}\tilde J_\s,
\qquad\qquad
\tilde J_\s=-\frac{1}{3} \eps_{\s\mu a b}{\cal S}^{\mu,ab}\,.
\label{dual-J5}
\ee

Therefore, we see that the spin current in the Dirac theory is
completely antisymmetric at classical level: it is the dual of the axial
current \cite{Hongo}.

It follows that the Dirac fermion in flat space only couples to the totally
antisymmetric part of the spin connection, that can be expressed in
terms of the axial background,
\be
\tilde A_\mu =-\frac{1}{2} \eps_{\mu\nu\r\s} \w^{\nu,\r\s},
\qquad\qquad 
\w_{\nu,\r\s} =\frac{1}{3} \eps_{\nu\r\s\l} \tilde A^\l +{\rm other\ terms}.
\label{spin-ax}
\ee

These relations can be used to `geometrize' the axial coupling.
We shall use this correspondence to find out how the
hydrodynamic theory couples to general backgrounds with torsion.

The relation between spin and axial currents \eqref{dual-J5}
extends to a curved space by including vierbeins to match
the index types, so it is a `duality' in a weaker sense\footnote{See 
eq. (2.13) of Ref.\cite{Hongo}.}
\be
    {\cal S}^{\mu,\nu\r}={\cal S}^{\mu,ab}E_a^\nu E_b^\r=
    \frac{1}{2}\epsilon^{\mu\nu\r\s}\tilde J_\s\,,
 \label{dual-J5-c}
\ee
where $E^\mu_a$ is the inverse vierbein. Note, however, that the spin
current is modified at the quantum level, since the contribution by
the axial-gravitational anomaly \eqref{S-cons} is not completely
antisymmetric.

We also compare the gauge transformations of both backgrounds
\be
    \d \w_{\mu,\r\s}=\left(\de_\mu\L -i[\L,\w_\mu]\right)_{\r\s}
    \sim \eps_{\mu\r\s\b} \d \tilde A^\b, \qquad\quad \d\tilde A= d\tilde \l\;.
\ee

Evaluating the rhs of $\d \w$ for antisymmetric spin connections
\eqref{spin-ax} one finds the relation
\be
\de^\b\tilde \l\sim \eps^{\b\mu\r\s}\de_\mu \L_{\r\s} +
2i \L^{\b\a}\tilde A_\a\,,
 \label{spin-ax-gauge}
\ee
showing that the axial Abelian symmetry is a subgroup of the non-Abelian
Lorentz symmetry. Thus, the correspondence between axial and spin
connections is one-to-one only within this subspace.

%-5.3.2-%%%%%%%%%%%%%%%%%%%%%%%%%%%%%%%%%%%%%%%
\subsubsection{Rewriting the axial background as a flat geometry with
torsion}

In this Section, the relation between axial and spin currents is shown
to be one-to-one in the limit of flat metric, $e^a_\mu=\d^a_\mu$ and
non-vanishing spin connection $\w_{ab}^\mu$, which is only possible in
Einstein-Cartan geometries with torsion.

In these cases, the metric and spin connection are independent variables,
but the connection remains metric compatible
\be
    g_{\mu\nu}=e^a_\mu e^b_\nu \eta_{ab}\,, \qquad
    E^\r_a e^a_\s=\d^\r_\s\,, 
    \qquad
    D_\mu e^a_\nu=\de_\mu e^a_\nu -\G^\r_{\mu\nu}e^a_\r +
    \w_{\mu b}^a e^b_\nu=0\,.
\ee
The affine connection acquires an antisymmetric part that identifies
the torsion tensor, ${\cal T}_{\mu\nu}^a$, a geometric quantity independent
of metric,
\be
    {\cal T}^\r_{\mu\nu}=\G^\r_{[\mu\nu]}=E^\r_a\left(
    \de_{[\mu}e^a_{\nu]}+\w_{[\mu b}^a e^b_{\nu]}\right)\,.
\ee
It turns out that the spin connection can be divided in two parts
\cite{Hughes-torsion}: the first one is that obtained in absence of
torsion, which was considered in Section \ref{sec:52}: it is now
denoted by an open dot on top, $\mathring\w_\mu^{ab}$; the second part
is called contorsion $K^{ab}_\mu$, a tensor related to torsion, as
follows
\begin{align}
  &\w_\mu^{ab}=\mathring\w^{ab}_\mu(e) +K_\mu^{ab}\,,
    \label{K-def}  \\
  &\de_{[\mu}e^a_{\nu]}+\mathring\w_{[\mu b}^a e^b_{\nu]}+
    K^{ab}_{[\mu}e_{b\nu]}={\cal T}^a_{\mu\nu}\,.
    \label{Tor-def}
\end{align}
We define the contorsion 1-form, $K^{ab}=K^{ab}_\mu dx^\mu$,  and
the torsion 2-form,
${\cal T}^a=\frac{1}{2}{\cal T}^a_{\mu\nu}dx^\mu\wedge dx^\nu$.
The Riemann 2-form also splits into
\be
    R^a_b=\mathring R^a_b + (\mathring D K)^a_b + K^a_c K^c_b\,,
 \label{R-def}
\ee
where $\mathring D_\mu$ is the covariant derivative with respect to
the Riemannian metric.

We assume a flat metric, $e^a_\mu=\d^a_\mu$,
$\mathring\w^a_{\mu b}=0$, and compute the geometry given by the spin
connection corresponding to the axial background, using the relations
\eqref{dual-J5} and \eqref{spin-ax} (neglecting `other terms'), i.e.
\be
    \w_{\nu,\r\s} =\frac{1}{3} \eps_{\nu\r\s\l} \tilde A^\l\,.
\ee
We thus compute the quantities
\begin{align}
  \w_{\mu,\nu\r}&\equiv\w_{\mu,ab}e^a_\nu e^b_\r
    = K_{\mu,\nu\r}=\frac{1}{3}\eps_{\mu\nu\r\l}\tilde A^\l\,,
  \\
    {\cal T}_{\a,\mu\nu}&\equiv {\cal T}_{a,\mu\nu}e^a_\a=2 \w_{\mu,\a\nu}\,,
  \\
    R_{\mu\nu,\a\b}&=(\de_{[\mu} K_{\nu]} +K_{[\mu}K_{\nu]})_{\a\b}=
    \frac{1}{3} \de_{[\mu}\eps_{\nu]\a\b\l}\tilde A^\l\,,
    \\
  4[ \Tr(R^2)]&=R_{\mu\nu,\a\b}R_{\r\s,\b\a}\eps^{\mu\nu\r\s}
=\frac{8}{9} [d\tilde Ad\tilde A]\,,
  \\
{\cal R}_{\a\b}&=R_{\mu\a,\mu\b}=-\frac{1}{3}\eps_{\a\b\mu\l}\de^\mu\tilde A^\l
       = -{\cal R}_{\b\a}\,.
\end{align}
We observe that:

- In the evaluation of the Riemann tensor, the term quadratic in $K$ vanishes.

- The Ricci tensor ${\cal R}_{\a\b}$ is totally antisymmetric, a
characteristic property of geometries with torsion, since $\cal R$ is
symmetric in Riemannian metric.

These results show how the axial field geometrizes into a torsion-only
background.

We now consider the hydrodynamic theory with $A=0,\, \tilde A\neq 0$
and flat metric and rewrite it in geometric form, i.e. in terms of the
spin current. Recall the expression of the action \eqref{S-bb-4d} and
the consistent axial current:
\begin{align}
    &S=\int d^4x\; P(p) +
    \int_{{\cal M}_4} (\tilde A-d\psi)pdp +\a\psi d\tilde A d\tilde A\,,
 \\
    & *\tilde J_{cons}=pdp +2\a d\psi d\tilde A\,, \qquad\qquad
    \de_\nu\tilde J^\nu_{cons}=-\a[d\tilde A d\tilde A]\,.
\end{align}
In this action, we replace the axial field with the
spin connection using previous formulas, and obtain
\be
    S=\int\; d^4x P(p) -
    \left (3 \w^{\mu,\nu\r} -\eps^{\mu\nu\r\s}\de_\s\psi \right)
    p_\mu\de_\nu p_\r
    -\a\frac{9}{8}\psi R_{\mu\nu,\a\b}R_{\r\s,\b\a}\eps^{\mu\nu\r\s}\,.
 \label{geo-act}
\ee
The spin current is obtained by variation of this action over
the spin connection,
\be
    {\cal S}^{\mu,\nu\r}=-\left\{  p^\mu\de^\nu p^\r  \right\}
    - 9\a\eps^{\mu\a\b\g}\de_\a\psi R_{\b\g}^{\ \ \ \nu\r}\,,
 \label{new-spin}
\ee
where braces represent complete antisymmetrization over the three indices,
\be
    \{a^\mu b^\nu c^\r\}=
    - \eps^{\mu\nu\r\s}\eps_{\s\mu'\nu'\r'} a^{\mu'}b^{\nu'} c^{\r'}\,.
\ee

The result \eqref{new-spin} shows that the spin current acquires a
totally antisymmetric classical term
$\left\{ p^\mu\de^\nu p^\r \right\}$ due to the coupling of torsion to
the 3-form $pdp$ in the action \eqref{geo-act}.  This coupling is not
possible in Riemannian backgrounds. There is a notorious ambiguity in
the definition of the spin current \cite{yarom-torsion}, in which one
could add $\left\{ p^\mu\de^\nu p^\r \right\}$ to the spin current
with an arbitrary coefficient. Without this addition the spin current
is entirely anomalous (the last term in \eqref{new-spin}), in
agreement with the results of Section \ref{sec:52}. The
relation \eqref{dual-J5} in the Dirac theory allows us to fix this
classical contribution to the spin current: while it is non-minimal,
its coefficient vanishes in the absence of torsion.

In the expression \eqref{new-spin}, the anomalous term is also
present, which is not fully antisymmetric, as already observed.
Actually, it is known that the $R^2$ anomaly \eqref{grav-ax} keeps the
same form in geometries with torsion \cite{Yajima:1985jd}.
The value of the anomaly
coefficient $\b$ is clearly not reproduced in the present setting
which `Abelianizes' the local Lorentz symmetry.

In conclusion, we have found that it is possible to geometrize the
axial background: it maps into torsion and can be described in the
setting of Einstein-Cartan gravity.  The spin current shows a
classical term which is not determined by anomaly inflow and vanishes
for Riemannian backgrounds.

%-5.3.3-%%%%%%%%%%%%%%%%%%%%%%%%%%%%%%%%%%%%%%%

\subsubsection{Spin current in  backgrounds with curvature and torsion}

We now discuss the general case of couplings to the $A,\, \tilde A$
backgrounds and gravity with torsion, which is described by the following
3+1d action
\begin{align}
    S[\pi,\psi] &= \int d^4x\,\sqrt{-g}\, P(p)
 \nonumber\\
    & +\int_{{\cal M}_4} \tilde{\cal A} (\pi -A)d(\pi+A) +
    \psi \left(d\pi d\pi +\alpha d\tilde A d\tilde A +\beta \Tr (R^2)\right)\,,
 \label{S-tor}
\end{align}
with
\be
    \tilde{\cal A}_\mu = \tilde A_\mu +
    \lambda \epsilon_{\mu\alpha\beta\gamma}\Gamma^{\alpha,\beta\gamma}
    = \tilde A_\mu + \frac{\lambda}{2}
    \epsilon_{\mu\alpha\beta\gamma}{\cal T}^{\alpha,\beta\gamma}\,.
 \label{A-tor}
\ee
The interaction with the completely antisymmetric connection is suggested
by the analysis of the previous Section, and is expressed in terms of
torsion ${\cal T}^\a_{\b\g}$ (see \eqref{Tor-def}).
The `generalized' axial background $\tilde{\cal A}$ allows for
fermionic excitations with arbitrary coupling to torsion
parameterized by the coefficient $\l$.

We remark that covariant derivatives of forms can involve
the torsion: e.g., for the one-form $a$ we have
\be
    D_{[\mu}a_{\nu]}=\de_{[\mu}a_{\nu]} - \G^\a_{[\mu\nu]}a_\a=
    \de_{[\mu}a_{\nu]} - {\cal T}^\a_{\mu\nu}a_\a\,.
 \label{tor-term}
\ee
The second term in the rhs of this equation is itself a tensor, thus
the first term involving ordinary derivatives is also covariant, being
the difference of two tensors. This implies that covariantization of
derivatives of forms is not compulsory: we can keep such terms in the
action \eqref{S-tor} unchanged, as a minimal choice.
Note also that the interaction between axial current and torsion
introduced via the $\tilde{\cal A}$ background \eqref{A-tor} is a
non-canonical coupling, which is physically motivated by the previous
study of the Dirac theory.

In the following we determine the currents and conservation laws
implied by the action \eqref{S-tor}. Hereafter we use covariant
derivatives with respect to the metric only, denoted by
$\mathring D_\mu$. We consider backgrounds with completely
antisymmetric contorsion, as those entering in \eqref{A-tor} (using
$\G_{\a,[\b\g]}={\cal T}_{\a,\b\g}=K_{[\b,\a\g]}$); we also denote
quantities in absence of torsion with a dot, as in the previous
Section.

Variations over $\pi$ and $\psi$ produce the following equations of motion
\begin{align}
  &\mathcal{J}^\nu \equiv -\frac{\p P}{\p p_\nu} +
    \Big[2\tilde{\cal A} d\pi -pd\tilde{\cal A}\Big]^\nu =
    \Big[2d\psi d\pi\Big]^\nu\,,
 \\
    &d\pi d\pi +\alpha d\tilde A d\tilde A +\beta \tr R^2 =0\,.
\end{align}
We obtain from these equations the transport consequences
\begin{align}
  &\mathcal{J}^\nu(\p_\nu\pi_\mu-\p_\mu \pi_\nu) =
    \mathring D_\mu\psi\Big[d\pi d\pi\Big]\,,
 \\
    &\mathring D_\nu \mathcal{J}^\nu =0\,,
 \\
    &\mathcal{J}^\nu \mathring D_\nu\psi =0\,.
\end{align}

The consistent currents are:
\begin{align}
  &J^\mu_{cons} = 
    -\frac{\p P}{\p \pi_\mu} +[(\pi- A)d\tilde A +2 \tilde A dA]^\mu \,,
 \\
  &\tilde J^\mu_{cons} =     [(\pi-A)d(\pi+A)+2\a d\psi d\tilde A]^\mu\,.
\end{align}
Using equations of motion, we find the conservation laws
\begin{align}
    \mathring D_\mu J^\mu_{cons} &= 0\,,
 \label{cons-div-101} \\
  \mathring D_\mu\tilde J_{cons} &=
 -[dAdA]-\a[d\tilde A d\tilde A] -\b [\Tr R^2]\,.
 \label{cons-div-102}
\end{align}
Note that they have the same form as in absence of torsion.

The consistent spin current is obtained by variation with
respect to the spin connection; 
\begin{align}
  {\cal S}_{\nu,\a\b}
  =&2\l\varepsilon_{\nu\a\b\s}[(\pi-A)d(\pi+A)]^\s
+2 \b g_{\nu\nu'}\eps^{\nu'\mu\a'\b'}\de_\mu \psi\; R_{\a'\b',\a\b}
  \\
  =& {\cal S}_{\nu,\a\b\; class} + {\cal S}_{\nu,\a\b\; cons} \,.
     \label{sigma-999}
\end{align}
Beside the anomalous term, the completely antisymmetric
classic term is also present owing to the linear dependence  of the
spin connection on contorsion and torsion in \eqref{K-def}.

We introduce the standard hydrodynamic stress tensor
\begin{align}
    T^\nu_{\ \ \mu} = -\frac{\p P}{\p \pi_\nu}(\pi-A)_\mu +P\delta^\nu_\mu \,,
  \label{T-end}
\end{align}
and derive its conservation law using the equations of motion\footnote{
  With the help of the 3+1d identity
  $p_{[\nu}\eps_{\r]\a\b\g}\G^{\a\b\g}=-3\eps_{\nu\r\a\b} p_\g\G^{\a\b\g}$,
  valid for totally antisymmetric affine connection.}; after some
calculations, we obtain:
\begin{align}
\mathring D_\nu T^\nu_{\ \ \mu} =&
F_{\mu\nu}J^\nu_{cov} +\tilde F_{\mu\nu}\tilde J^\nu_{cov}
+3\l \Big[(\pi-A)d(\pi+A)\Big]^\nu\eps_{\nu\mu\a\b} \mathring D_\g\G^{\a,\b\g} 
\nonumber\\
&-\beta(\p_\mu\psi -\tilde{\cal A}_\mu)\Big[\Tr (R^2)\Big]\,.
    \label{T-cons-tor}
\end{align}
In this expression, there appear the following generalized
currents and their anomalies
\begin{align}
  J^\nu_{cov} &= -\frac{\p P}{\p \pi_\nu} +
                \Big[(\pi-A)d\tilde{\cal A}\Big]^\nu\,,
\label{J-end} \\
  \tilde J^\nu_{cov} &= \Big[(\pi-A)d(\pi+A)+
                       2\a (d\psi-\tilde{\cal A})d\tilde A\Big]^\nu\,,
\label{Jt-end} \\
    \mathring D_\nu J^\nu_{cov} &= -\Big[2d\tilde{\cal A}dA\Big]\,,
\label{J-cons-end} \\
  \mathring D_\nu \tilde J^\nu_{cov} &=
         -\Big[dAdA+2\alpha d\tilde A d\tilde{\cal A}
  +\alpha d\tilde A d\tilde A +\beta \Tr (R^2)\Big]\,.
\label{Jt-cons-end}\end{align}
Note that the covariant anomalies are consistent with the inflow from
the generalized 4+1d response action
\be
    S= - \int_{{\cal M}_5} \tilde{\cal A}\left( dA dA +\a d\tilde A d\tilde A +
    \b  \Tr(R^2) \right)\,.
\ee
In this expression, the field $\tilde{\cal A}$ has been generalized in
4+1 dimensions by introducing the fully antisymmetric tensor
$\hat{\cal T}^{a\b\g\s}$, whose component $\s=5$ is only nonvanishing
at the boundary, where it matches the torsion
$\hat{\cal T}^{a\b\g 5}={\cal T}^{a\b\g}$.

The right-hand side of the stress tensor conservation
\eqref{T-cons-tor} can be rewritten as follows.
The fourth term is identified as the contribution due to the
anomalous spin current ${\cal S}_{cov}$ as in \eqref{T-113}. Following 
the same steps as in \eqref{cons-rev}, we find,
\be
    -\beta(\p_\mu\psi -\tilde{\cal A}_\mu)\Big[\Tr (R^2)\Big]=
    2\b\Tr(R_{\mu\s}[(d\psi-\tilde{\cal A})R]^\s)=
    \frac{1}{2}\Tr (R_{\mu\s} {\cal S}^\s_{cov})\,.
\ee
We can include the corresponding term with the
classical spin current which can be written
\be
    \frac{1}{2}\Tr(R_{\mu\nu}{\cal S}^\nu_{cl})=\frac{1}{2}
    \Tr\left((\mathring R_{\mu\nu}+ \mathring D_{[\mu}K_{\nu]}
    +K_{[\mu}K_{\nu]} ){\cal S}^\nu_{cl}\right)=\frac{1}{2}
    \Tr(\mathring D_{[\mu}K_{\nu]}{\cal S}^\nu_{class})\,,
\ee
by observing that the two of the three terms of the Riemann
curvature (cf. \eqref{R-def}) vanish, the first once contracted
with the totally antisymmetric $S_{cl}$ owing to the second Ricci identity,
while the third one by full antisymmetry of contorsion.

The third term in the r.h.s. of \eqref{T-cons-tor} involves the classical spin current contracted with torsion
\be
    3\l \Big[(\pi-A)d(\pi+A)\Big]^\nu\eps_{\nu\mu\a\b} \mathring D_\g\G^{\a,\b\g}
    =\frac{3}{2} {\cal S}_{\mu,\a\b\; cl}\mathring D_\g {\cal T}^{\g,\a\b}\,.
\ee

We finally recast the stress-tensor conservation in the following form
\begin{align}
    \mathring D_\nu T^\nu_{\ \mu} &=
    F_{\mu\nu}J^\nu_{cov} +\tilde F_{\mu\nu}\tilde J^\nu_{cov}
    +\frac{1}{2}\Tr (R_{\mu\s} {\cal S}^\s_{tot})
 \nonumber\\
    &-3\mathring D_\g K^{\g,\a\b} {\cal S}_{\mu,\a\b \; class}
    +\frac{1}{2}\mathring D_{[\mu} K_{\s],\a\b}{\cal S}^{\s,\a\b}_{class}\,,
 \label{T-cons-end}
\end{align}
where ${\cal S}_{tot}= {\cal S}_{class} + {\cal S}_{cov}$.  We observe
that for vanishing torsion this conservation law agrees with the
expression found in Section \ref{sec:52}, there obtained by the
diffeomorphism Ward identity. This provides a consistency check for
our approach. Note also that the spin-current term \eqref{T-cons-end}
is cubic in curvature and torsion, owing to the non-minimal coupling
introduced in \eqref{A-tor}.

In conclusion, we obtained the so-called constitutive relations for
currents \eqref{J-end}, \eqref{Jt-end} and stress tensor \eqref{T-end}
and respective conservation equations \eqref{J-cons-end},
\eqref{Jt-cons-end} and \eqref{T-cons-end} in most general
backgrounds.

% -6-%%%%%%%%%%%%%%%%%%%%%%%%%%%%%%%%%%%%%%%

\section{Physical aspects of the variable $\psi$}
\label{sec:6}

The description of anomalous 3+1d hydrodynamics and the matching with
4+1d topological theories has led to the appearance of a new
pseudoscalar field $\psi$, which ensures gauge invariance of the
axial quantity $\tilde q=d\psi -\tilde A$.  In the following we
suggest some physical effects associated to this field.

%-6.1-%%%%%%%%%%%%%%%%%%%%%%%%%%%%%%%%%%%%%%%
\subsection{Spinor rotation}
\label{sec:61}

The relation between axial and spin currents in the Dirac theory
described in Section \ref{sec:531} provides some geometric insight.
We have seen that the axial gauge transformation associated to $\psi$
maps into a local Lorentz transformation
(cf. \eqref{spin-ax-gauge}). Specifically, for a gauge field in the
third direction $\tilde A_3$, one finds
\be
    \de_3\tilde\l \sim \dot \L_{12}\,,\qquad\qquad \psi\ \to\ \psi+ \tilde\l\,,
\ee
where the dot indicates a time derivative. The gauge transformation
$\L_{12}$ is absorbed by a rotation of Dirac spinors in the direction
$\hat 3$, orthogonal to the $(12)$ plane.  This correspondence
suggests that the variable $\psi$ is associated to spin degrees of
freedom of the Dirac fluid.

%-6.2-%%%%%%%%%%%%%%%%%%%%%%%%%%%%%%%%%%%%%%%
\subsection{Interpretation as dynamic chiral chemical potential}

In some physical settings, the axial background is used to mimic
external conditions that cause an imbalance between the two
chiralities of fermions
\cite{fukushima2008chiral,arouca2022quantum}. In particular, the axial
potential can be seen as the difference of chemical potentials for
left and right fermions,
\be
    \tilde A_0=\mu_R -\mu_L\,.
 \label{mu-umbal}
\ee
In the following, we shall argue that the same interpretation can be
given to the companion quantity $\dot\psi$, which is a dynamic degree
of freedom of the fluid.

Let us illustrate this feature by considering the stress tensor
conservation \eqref{T-102} in presence of vector background
$A\neq 0$ and $\tilde A=0$,
\begin{align}
    &  \de_\nu T^\nu_\mu=F_{\mu\a}J^\a_{cov}\,,
 \label{T-chem}\\
    &J_{cov}^\a=J^\a_{cons}=2[d\psi d\pi]^\a\,,
\end{align}
where we used \eqref{J-bad} for the expression of the covariant
current evaluated on the equations of motion. We write \eqref{T-chem}
for $\mu=0$, expressing the adiabatic change of energy in the fields
$(\vec E, \vec B)$, assumed static or slowly varying,
\be
    \dot {\cal E}=2 F_{0i} \eps^{i0jk}\dot \psi\de_j(p_k+ A_k)=
    2 \vec E \cdot \vec B\; \dot{\psi}\,,
 \label{E-change}
\ee
where the fluid velocity is taken to vanish ($p_k\sim v_k=0$, $ k=1,2,3$).

The result \eqref{E-change} should be compared with the expression
describing the spectral flow\footnote{See, e.g., Section 4.4 of
  \cite{arouca2022quantum} for a detailed discussion of spectral flow
  in 3+1 dimensions.}, a characteristic feature of chiral anomalies,
which is also realized in the same backgrounds for small
$\dot{\vec E}$:
\be
    \dot Q_L-\dot Q_R =2\; {\rm Vol}\; \vec E \cdot \vec B\,,
 \label{Q-change}
\ee
where Vol is the volume of the system.

The difference between the two physical settings is the following. In
the spectral flow \eqref{Q-change}, the charge non-conservation is due
to left (resp. right) particles moving above (below) the Fermi level,
without energy change.  The same effect can be obtained by an
imbalance of chemical potentials \eqref{mu-umbal}.

In the system described by \eqref{E-change}, there is energy variation
$\dot {\cal E}\propto \dot \psi$ in the same gauge background.  We
interpret this effect as an imbalance of chemical potential between
left and right particles, that adds to the spectral flow, causing the
energy change. Actually, the equation \eqref{mu-umbal} should be
modified owing to axial gauge symmetry into
\be
    \tilde A_0-\dot \psi =\mu_R -\mu_L\,,
\ee
supporting the interpretation that $\psi$, solution of the equations
of motion, can realize a dynamic chiral imbalance also in absence of
external background $\tilde A$.

%-6.3-%%%%%%%%%%%%%%%%%%%%%%%%%%%%%%%%%%%%%%%

\subsection{Adding dynamics to $\psi$}
\label{sec:63}

We have seen that the action and equations of motion depend on the
axial gauge invariant quantity $\tilde q = d\psi-\tilde A$. While we
consider $\tilde A$ as a background field, the $\psi$ field
corresponds to a physical degree of freedom. Fixing the gauge for
$\tilde A$ and solving equations of motion for $\tilde q$ we can
determine the corresponding configuration of $\psi$. We remark that
$\psi$ might be thought of as an axial superfluid phase.

Let us first recall the case of sigma models, low-energy approximation
of gauge theories, where gauge degree of freedoms become physical and
acquire a dynamics. In the Abelian Higgs model,  the gauge
parameter for vector gauge symmetry is $\th(x)$.
The action obtained by writing
the complex relativistic scalar field as $\phi=\r\exp(i\th)$ reads
\cite{weinberg1986}
\be
S=\int F(A_\mu -\de_\mu \th) +\r^2(\de_\mu\th -A_\mu)^2 +
(\de_\mu\r)^2 +V(\r)\,.
\ee
The last three terms express the dynamics for $\th$:
in presence of spontaneous symmetry breaking, 
$<\r^2>\neq 0$, $\th$ acquires a $\s$-model dynamics,
to which we can add WZW terms for anomalies.
Note that other phases are possible with $<\r^2>= 0$.

We think that the $\psi$ dependence and its dynamics could be
explicitly found by integrating out e.g. the free fermion theory in
presence of fixed currents. The proof of this fact requires
independent studies and is not given here.

Let us nonetheless check that adding a dynamics for $\psi$ does
not affect the form of anomalies. We already know that
these are independent of the form of the pressure $P(\pi-A)$,
 in the action \eqref{S-bb-4d}, including the case $P=0$.
 We can add the $\psi$ dependence to pressure in a way
 that respects gauge invariance, as follows,
\be
P=P(\pi-A, d\psi-\tilde A), \qquad\qquad
(p=\pi-A,\ \tilde q=d\psi -\tilde A)\,.
\label{P-gener}
\ee
The equation of motion for $\psi$ \eqref{eqm-4d-con} and the
axial current \eqref{jtcons-al} are modified as follows:
\begin{align}
  -\partial_\nu \frac{\de P}{\de \tilde q_\nu}+&
  \left[d\pi d\pi +\a d\tilde A d\tilde A \right]=0\,,
  \\
  \tilde J^\nu_{cons}=&- \frac{\de P}{\de \tilde q_\nu} +
  \left[(\pi-a )d(\pi +A)+2\a d\psi d\tilde A\right]^\nu\,.
\end{align}

One verifies that the anomaly \eqref{jt-anom-corr} is unchanged
\be
\de_\nu \tilde J^\nu_{cons}= -\de_\nu \frac{\de Q}{\de \tilde q_\nu} +
\left[d\pi d\pi -dA dA\right]
\underset{eq.m.}{=} -\left[ dA dA +\a d\tilde A d\tilde A \right]\,.
\ee
This is clearly a simple consequence of axial gauge invariance, i.e. of
the functional form $d\psi -\tilde A$.

Depending on the phase of matter for the fluid under consideration the
equation of state might or might not depend on
$d\psi -\tilde A$. The stiffness with respect to $d\psi-\tilde A$
would mean a presence of an axial superfluid condensate. Contrary, if
the only dependence on $d\psi-\tilde A$ comes in the form of
gradients, the corresponding phase of matter would be characterized by
a weakly fluctuating axial charge.

%-6.4-%%%%%%%%%%%%%%%%%%%%%%%%%%%%%%%%%%%%%%%

\subsection{Chiral symmetry breaking} 

Let us consider a pressure function \eqref{P-gener} that is not axial
invariant and depends explicitly on $\psi$, namely
$P=P(d\th -A, \psi )$. Repeating the previous steps, we now obtain
\be
\de_\mu \tilde J^\mu = -\left[ dA dA +\a d\tilde A d\tilde A \right]
- \frac{\de P}{\de\psi}\,.
\label{O32}
\ee
The additional classical term is explicitly breaking the symmetry: due to
its definition, it is a pseudoscalar quantity and vector gauge invariant,
thus it can be interpreted as
the mass term $m\langle \bar\Psi \g^5 \Psi\rangle$ in Dirac theory
or its generalization in interacting cases.

%-7-%%%%%%%%%%%%%%%%%%%%%%%%%%%%%%%%%%%%%%%%
\section{Conclusions}

In this work, we developed the Euler hydrodynamic description of 3+1
dimensional Dirac fermions which is based on the action \eqref{S-bb-4d},
reproduced here for convenience,
\begin{align}
    S_I=  &- \int_ {{\cal M}_5} \tilde AdAdA+\a \tilde Ad\tilde A d\tilde A
\nonumber\\
      &+\int_{{\cal M}_4} P\left(\pi-A\right) +
        \tilde A(\pi-A)d(\pi+A) +
    \psi(d\pi d\pi+\a d\tilde A d\tilde A)\,.
 \label{S-minim}
\end{align}
and the free variation with respect to the fields $(\pi,\psi)$.
This theory describes the axial anomalies in $A, \tilde A$ backgrounds
and, with the additions in Section \ref{sec:5}, the mixed axial-gravitational
anomaly in geometries with curvature and torsion.
Rather general dynamics, i.e. equations of state,
of the fluid are encoded in the form of the pressure
function $P$, which may also depend on the other gauge invariant
quantity $d\psi -\tilde A$, as explained in Section \ref{sec:63}.
This effective theory description demonstrates explicitly that anomalies are 
independent of dynamics, namely of the form of the pressure.

Our approach makes it clear that the topological theory in one extra
dimension provides the main inputs for defining the hydrodynamics:
through the anomaly inflow relation between bulk and boundary
currents, it establishes the fluid degrees of freedom entering the
description and the anomaly content.

In Section \ref{sec:43} we found two further topological theories
which reproduce the same Dirac response action \eqref{S5-resp} and
anomalies.  These theories can model hydrodynamics featuring
additional degrees of freedom: the action (\ref{S_II}) involves
independent vector and axial momenta, and the expression (\ref{action-90})
presents currents parameterized 
by two-form fields.  Other hydrodynamics might exist involving
additional extended excitations expressed by higher-form fields and
their generalized symmetries \cite{Gaiotto-higher}.
The coupling to higher-form background fields
should also be considered because it generates corresponding
anomalies which help characterizing the theories.

Another possible extension of this work involves the description of fermionic
theories with different anomaly content, as e.g. with multi-component
Abelian and non-Abelian symmetries \cite{jackiw-rev,nair:2013}.

These results provide an interesting starting point for developing
effective field theories for bosonization of 3+1 dimensional
fermions. The rather simple variational formulation of hydrodynamics
described in this work is very close to bosonic effective field
theory, actually identical to it in 1+1 dimensions.  Both approaches
are semiclassical, low-energy descriptions, only valid for local
quantities and response functions.  Yet, hydrodynamics can capture
non-perturbative phenomena in many physical systems
\cite{kharzeev2011testing}.

The action \eqref{S-minim} defines an effective theory for fluid
phases of interacting massless fermions.  Note that the pressure term
in \eqref{S-minim} takes the form
$P=P(\mu^2)=P\left(-(\pi-A)_\mu^2\right)$ in the relativistic case,
and it cannot be immediately identified as a field-theory kinetic
term. Solving explicitly the constrains proportional to $\psi$ in the
action \eqref{S-minim} might bring its expression into a more
familiar form.

The extension of our approach to the chiral case of a single Weyl
fermion is also challenging: the two-fluid theory \eqref{S_II} is a
good starting point, by splitting it into two equal parts. Our current
understanding is that the Weyl theory should involve fluid
variables with 4+1 dimensional terms in the action, which cannot be
reduced to 3+1 dimensions as realized in \eqref{S-minim}.  A related
proposal is to describe Weyl fluids by strong chiral unbalancing of
Dirac fluids \cite{Wiegmann-Weyl}.

Finally, a natural program is that of introducing temperature and
entropy in hydrodynamics. This can be done within the same variational
approach.

Our methods are closely related to other variational approaches
for so-called adiabatic fluid dynamics with
anomalies, as  those constructed in
\cite{Jensen-transgr,Haehl:2013hoa,haehl2015adiabatic} for example.
The main difference is in the type of
independent fields used in the action functional. Here we considered the
canonical momentum per particle and scalar fields playing the role of
Lagrange multipliers for the constraints of the theory. This choice
made the discussion closely related to bosonization of
one-dimensional fermions. We generally view the obtained actions for
fluid dynamics as effective bosonic nonlinear theories of strongly
coupled fermionic fluids.

\bigskip

%%%%%%%%%%%%%%%%%%%%%%%%%
%%%%%%%%%%%%%%%%%%%%%%%%%

\acknowledgments

We gratefully acknowledge P. Wiegmann for collaboration in the early
stages of this work and for useful criticism. We also benefit from
interesting exchanges with J. Fr\"ohlich, F. M. Haehl, K. Jensen,
B. Khesin, N. Nekrasov, N. Pinzani-Fokeeva, R. Villa, A. Yarom.
A.C. would like to thank Marco Ademollo, Marcello Ciafaloni and
Stefano Catani for sharing their passion for theoretical physics.  The
work of A.C. has been partially supported by the grant PRIN 2017
provided by the Italian Ministery of University and Research. The work
of A.G.A. has been supported by the National Science Foundation under
Grant NSF DMR-2116767 and by NSF-BSF grants 2020765, 2022110.
A.G.A. also acknowledges the support of Rosi and Max Varon fellowship
from the Weizmann Institute of Science and the hospitality of the
Galileo Galilei Institute for Theoretical Physics.

\appendix

%-A-%%%%%%%%%%%%%%%%%%%%%%
\section{Variational principles for perfect barotropic fluids}
\label{sec:litvar}

\subsection{Introduction and constrained variations}

 We start by summarizing some basic facts about Euler hydrodynamics
 \cite{jackiw-rev}.  The fluid is described, in the
 non-relativistic case, by the density and velocity fields
 $(\r,v^i)$ parameterizing the current, $J^\mu=(\r,\r v^i)$.  These
 obey the Euler and continuity equations,
\begin{align}
    & \de_0 v_i +v^k\de_k v_i= -\frac{1}{\r} \de_i P= - \de_i \mu\,,
 \label{e-eq}  \\
    &  \de_\mu J^\mu= \de_0 \r + \de_i (\r v^i)=0\,.
 \label{J-eq1}
\end{align}
We consider the limit of vanishing temperature, without entropy and
heat flows. In barotropic fluids the pressure depends only on density,
$P=P(\r)$, and obeys the thermodynamic relation $dP=\r d\mu$
in terms of the chemical potential $\mu(\r)$, which is also equal
to the enthalpy $h$ per particle in absence of heat flow.

We now briefly describe the constraints occurring in Euler dynamics.
In 1+1 dimensions, the Euler equation can be rewritten as
\begin{align}
\de_t v = \de_x \left( -\mu -\frac{v^2}{2} \right)\,, \qquad {\rm i.e.} \qquad
                \de_0 p_1 =\de_1 p_0\,,   \qquad   
 (p_0,p_1)= \left( -\mu -\frac{v^2}{2}, v \right)\,.
\end{align}
This is the constraint\footnote{In this Appendix we consider vanishing
  gauge backgrounds, thus the fluid momentum is $\pi\equiv p$.} $dp=0$
discussed in Section \ref{sec-2}: it implies the following invariant
of motion
\be
    C_1=\int dx\; v\,, \qquad\qquad \de_0 C_1=0\,.
\ee

In 3+1 dimension the Euler equation implies the following relation
obeyed by the fluid vorticity, $\bm \w = \bm \nabla \times \bm v$, 
\be
\de_0\bm \w=\bm\nabla\times(\bm v \times \bm\w)\,,
\qquad\qquad \qquad\qquad
 \bm \w = \bm \nabla \times \bm v\,.
\label{w-eq}
\ee
It is also instructive to rewrite the Euler equation \eqref{e-eq} in
four-dimensional covariant form by using the four-current 
and introducing the fluid four-momentum
\be
p_\mu=\left(-\mu - \frac{v^2}{2}, v^i\right)\,.
\label{p-def}
\ee
We find,
\be
\p_0p_i-\p_ip_0 +v^j(\p_jp_i-\p_ip_j) =0\,,
\qquad {\rm i.e.}\qquad J^\mu (\p_\mu p_\nu-\p_\nu p_\mu)=0\,.
 \label{e-cov}
 \ee
 
In 3+1 dimension, the constraint $dpdp=0$ discussed in Section
\ref{sec:1+1hydro} amounts to the conservation of the helicity current
  $\tilde J= *(pdp) $. The associated charge is
\be
C_3=\int d^3x\; v^i \w_i\,, \qquad\qquad \de_0 C_3=0\,.
\ee
The time independence of this quantity can be proven by using the
Euler and vorticity equations \eqref{e-eq}, \eqref{w-eq} or 
more straightforwardly from (\ref{e-cov}).

These results show that the action variational principle for the Euler
equation \eqref{e-eq} cannot be formulated in terms of the original
variables $(\r, \bm v)$, because variations should respect the constraints.
As discussed in the main text, one solution is to vary with respect to
the Clebsch scalar parameters which locally solve the constraints
\be
1+1d:\ dp=0 \ \to \ p=d\th\,,\qquad\qquad 3+1d:\ dpdp=0\ \to \
p=d\th + \a d\b\,.
\ee
We refer the reader to
\cite{zakharov1997hamiltonian,jackiw-rev} for an introduction
to the formalism involving Clebsch parameterization of hydrodynamic
fields.

Another method described in Section \ref{sec:1+1hydro} is based on the use of
admissible (diffeomorphism) variations which leads to the
Carter-Lichnerowicz equations of motion reproducing the constraints.
This will be further analyzed in the following.

We first give the form of the action.  Both relativistic and
non-relativistic Lagrangians of a perfect fluid have been considered
\cite{schutz1970perfect,seliger1968variational}. In fact, one could
use a general variational formulation that is applicable independently
of the spacetime symmetry \cite{jackiw-rev}.  The action is
\begin{align}
  &S[J^\mu,p_\mu] = -\int d^4x\, \left[J^\mu p_\mu +
    \varepsilon(J^\mu)\right]\,.
 \label{act-1000}
\end{align}
It is a functional of the current $J^\mu$ and canonical momentum
per particle $p_\mu$.
The spacetime symmetries and the dynamics (equation of state) are
encoded in the scalar function
$\varepsilon(J^\mu)$ which is the energy density.

Even more compact formulation can be obtained by eliminating $J^\mu$ 
by a Legendre transform of the action,
leading to
\be
    S[p_\mu] = \int d^4x\, P(p_\mu)\,,
  \label{act-1002}
\ee
with
\be
   J^\mu = - \frac{\d S}{\d p_\mu}= -\frac{\de P}{\de p_\mu}\,.
\label{J-def}
\ee
The action is given by the spacetime integral of the pressure 
$P=P(\mu(p_\mu))$ which is considered as a function of the
canonical momentum $p_\mu$. Non-relativistic
and relativistic fluids differ in the expression for $\mu(p_\mu)$, 
 respecting the corresponding Galilean and Lorentz symmetries.
In the non-relativistic case, we have (cf. \eqref{p-def})
\begin{align}
     \mu =-p_0 -\frac{1}{2}p_ip^i, \qquad p^i=v^i\,,
\label{nr-param}
\end{align}
indeed reproducing the earlier form of the current
\begin{align}
  J^0 = -\frac{\p P}{\p p_0} = \frac{\p P}{\p \mu}=\r\,,
  \qquad J^i = -\frac{\p P}{\p p_i} = \frac{\p P}{\p \mu}p^i=\r v^i\,.
\label{J-form}
\end{align}
Again $\mu$ is understood as a function of $\rho$ and can be
obtained from the thermodynamic relations $dP =\rho d\mu$,
$\epsilon +P =\rho \mu$, etc.

In the main text we considered the action \eqref{act-1002} and 
use the admissible  (i.e. diffeomorphisms) variations
of $p_\mu$ to obtained the
Carter-Lichnerowicz equations of motion
\be
    J^\mu(\de_\mu p_\nu -\de_\nu p_\mu)=0\,,\qquad\qquad
    J^\mu\equiv-\frac{\de P}{\de p_\mu}\,,
 \label{CL-eq}
\ee
and current conservation.
The first equation in \eqref{CL-eq} is recognized as the Euler
equation in covariant form \eqref{e-cov}. This also imply the conservation of
the stress tensor
\begin{align}
  \de_\nu T^\nu_\mu=0\,,\qquad\qquad
  T^\nu_\mu &= J^\nu p_\mu +P\delta^\nu_\mu\,.
 \label{Tcr-1000}
\end{align}
These equations can be rewritten using \eqref{p-def} and
\eqref{J-eq1} as follows,
\begin{align}
  & \p_t \Big(\epsilon+\frac{\rho v^2}{2}\Big) +
    \p_i\left(\rho v^i \left(\mu +\frac{v^2}{2}\right)\right) = 0\,,
 \label{pE-1005}\\
    &\p_t (\rho v_i) +\p_j\Big(\rho v_i v^j +P\delta^j_i\Big) = 0\,.
 \label{pMom-1005}
\end{align}
The first and second equations are, respectively, the components $\mu=0$ and
$\mu=j$ of (\ref{Tcr-1000}): they express the local conservations
of energy and momentum.
The Euler equation \eqref{e-eq} is also recovered from the momentum 
\eqref{pMom-1005} and current \eqref{J-eq1} conservations.
We remark that among all equations described here
only four are independent. For example, the conservation of
energy and momentum (\ref{pE-1005}, \ref{pMom-1005}) can be derived
from the Euler and continuity equations \eqref{e-eq}, \eqref{J-eq1}
and vice versa.

%-A.2-%%%%%%%%%%%%%%%%%%%%%%
\subsection{Relativistic fluids}
 \label{sec:relf}

In this case the pressure depends
on the Lorentz invariant combination of momenta $p_\mu$:
\begin{align}
    P = P(\mu)\,, \qquad \mu =\sqrt{-p_\mu p^\mu}\,.
\end{align}
We introduce the 4-velocity $u_\mu$ as 
\begin{align}
    p_\mu = \mu u_\mu\,, \qquad u_\mu u^\mu =-1\,,
\end{align}
and the particle density $n$ as
\begin{align}
    n = \frac{\p P}{\p \mu}\,,
 \label{nPmu-1000}
\end{align}
so that the current is
\begin{align}
    J^\mu = -\frac{\p P}{\p p_\mu} = nu^\mu\,.
\end{align}
Upon using these parameterizations, the
stress tensor \eqref{Tcr-1000} and current \eqref{J-eq1} conservations
now imply the hydrodynamic equations for perfect relativistic
fluids \cite{landau6}
\begin{align}
    &\p_\nu\Big(\mu n\, u_\mu u^\nu +P\delta^\nu_\mu\Big) = 0\,,
 \label{pT-1005}\\
    &\p_\mu(nu^\mu) =0\,.
 \label{pJ-1005}
\end{align}
For barotropic fluid the chemical potential $\mu$ and pressure $P$ can
be considered as functions of the particle density $n$, obeying
the thermodynamic relations following from (\ref{nPmu-1000})
\begin{align}
    dP = n d\mu\,,\qquad d\epsilon = \mu dn\,, 
\end{align}
where $\epsilon(n)$ is the internal energy of the fluid related to the
pressure by Gibbs-Durhem formula
\begin{align}
    \epsilon + P = \mu n\,.
\end{align}

%-A.2-%%%%%%%%%%%%%%%%%%%%%%
\subsection{Free variations with constraint}
\label{sec:nrelf}

We now discuss in more detail the variational scheme adopted in this
paper, which is based on taking free variations while imposing the constraint
into the action via a Lagrange multiplier. We have
\begin{align}
    &S[p_\mu,\psi] = \int d^4x\, P(p_\mu) +\int_{{\cal M}_4} \psi\, dp dp\,. 
 \label{act-1004}
\end{align}
The second term is written as an integral of the 4-form to emphasize
its topological nature (independence on the metric). The equations of
motion obtained by free variations over $p_\mu$ and $\psi$ read:
\begin{align}
  &p_\mu: \qquad J^\mu\equiv -\frac{\p P}{\p p_\mu}=
    2 [d\psi dp]^\mu \,,
 \label{pi-1000}\\
    &\psi\ :\qquad  dp dp = 0\,.  
 \label{dpidpi-1000}
\end{align}
These expressions form a complete system
of equations needed to find the evolution of fields
$(p_\mu,\psi)$. As fully described in the main text, these imply the
stress tensor and current conservations, \eqref{Tcr-1000}
and \eqref{J-eq1}, and thus all earlier equations of Euler hydrodynamics in
both non-relativistic and relativistic cases.

We now discuss the equivalence between:
$(F)$ free-variation equations \eqref{pi-1000}, \eqref{dpidpi-1000} and 
$(CL)$ the Carter-Lichnerowicz  and conservation
 equations, \eqref{CL-eq}, \eqref{J-eq1}.

$ (F) \ \to\ (CL)$: Current conservation clearly follows from \eqref{pi-1000}.
Upon substituting this equation into \eqref{CL-eq}, we find
\be
J^\mu(\de_\mu p_\nu -\de_\nu p_\mu)=
2[d\psi dp]^\mu(\de_\mu p_\nu -\de_\nu p_\mu)=
-\de_\nu\psi [dpdp]=0\,,
\ee
Here we used the following 4d algebraic identity, also
employed in the main text,
\be
a_\mu [BC]=[aB]^\nu C_{\nu\mu} +[aC]^\nu B_{\nu\mu}\,,
\label{4d-id}
\ee
valid for any 1-form $a$ and 2-forms $B,C$. It is proven by checking the
vanishing of a 5-form in four dimensions.

$(CL) \ \to (F)$: The result depends on the rank of the antisymmetric
matrix $dp$, which can be zero or two, due to \eqref{CL-eq};
as explained in Section \ref{sec:311}, this implies that
 \eqref{dpidpi-1000} holds.
If the rank of $dp$ is two, we can choose local coordinates in which $dp$ is
non-vanishing along directions $x^2,x^3$, i.e.
$dp=\w_{23}dx^2 dx^3$. Then the current $J^\mu=-\de P/\de p_\mu$
has components along $x^0,x^1$; its dual 3-form is
\be
*J=(\eta dx^0 +\s dx^1)dx^2 dx^3\,,
\ee
where $\eta,\s$ are scalar functions. Current conservation \eqref{J-eq1}
implies $\de_1\eta -\de_0\s=0$, which can be integrated locally to
$*J=(\de_1\g dx^0-\de_0 \g dx^1) dx^2dx^3$. This is equation \eqref{pi-1000}
upon identification of $\psi=\g/\w_{23}$.

For vanishing rank $dp=0$, the constraint \eqref{dpidpi-1000} has a
double zero. The CL equation \eqref{CL-eq} is trivially satisfied and
\eqref{pi-1000} is not implied. The current $J^\mu$ is nonetheless
conserved and mildly constrained by $dp=0$ through the relation
between $p$ and $J$.  Equation \eqref{pi-1000} is instead different
because it forces the current to vanish.  This result is not
acceptable, because there can be locally non-vanishing fluid motion
($J\neq 0$) even in the absence of vorticity, e.g., in regions outside
of localized vortex lines, where $dp=0$.

We can resolve this conflict by technically assuming that $dp$ can
take very small values but never vanish. When it is small we can
allow the quantity $d\psi$ to grow large so as to have a finite
value of the current in \eqref{pi-1000}. The $\psi$ singularity is
`hidden' because this quantity disappears once equations are rewritten
in terms of original variables $(\r, \bm v)$.  Thus, this can be
viewed as a problem of coordinate singularities of the free
variation approach, not a real physical problem.

Furthermore, the issue is of limited interest in the generic 
case of both
non-vanishing backgrounds, because the equations of motion
  $d\pi d\pi +\a d\tilde A d\tilde A=0$ implies that $d\pi=0$ might
  occur only at points where $d\tilde A d\tilde A=0$. 
This is analyzed in the following subsection.
Further related aspects are discussed in two other Sections.

%-A.3-%%%%%%%%%%%%%%%%%%%%%%
\subsection{Variational principle in presence of
  nontrivial backgrounds}
 \label{sec:back}

Let us consider the following modification of \eqref{act-1004}
\begin{align}
  &S[p_\mu,\psi] = \int d^4x\, P(p_\mu) +\int_{{\cal M}_4} \psi\,
    (dp dp+dC)\,, 
 \label{act-1104}
\end{align}
where $C\neq 0$ is a non-trivial background 3-form field. Let us show that
the equations obtained from the free variational principle applied to
\eqref{act-1104} are identical to the ones obtained using restricted
variations common in the hydrodynamic literature.

The free variations give
\begin{align}
  {\mathcal J}^\nu \equiv &-\frac{\p P}{\p p_\nu}  =
                            2\Big[d\psi dp\Big]^\nu \,,
 \label{Jdp-1000}\\
    & dpdp +dC =0\,.
 \label{dpdp-1000}
\end{align}

On the other hand, we can derive equations of motion using the following
set of allowed variations
\begin{itemize}
    \item $\delta \pi = d\lambda\,, \qquad \delta \psi=0\,,$
    
    \item $\delta \pi =0\,, \qquad\ \ \delta \psi =\tilde \lambda\,,$
    
    \item  $\d \pi={\cal L}_\eps \pi\,, \quad \d \psi={\cal L}_\eps \psi\,,$
\end{itemize}
respectively corresponding to vector and axial gauge and diffeomorphism
transformations.
Using these variations parameterized by $\lambda,\tilde\lambda,\eps^\mu$
after some algebra we derive the following equations
\begin{align}
    &\p_\mu \mathcal{J}^\mu =0\,,
 \label{dJ-1100}\\
    & dpdp +dC =0\,,
 \label{dpdC-1100}\\
    &\left(\mathcal{J}^\nu -2\Big[d\psi dp\Big]^\nu\right)
      (\p_\nu p_\mu-\p_\mu p_\nu) = 0\,.
 \label{Jdp-1100}
\end{align}

It is obvious that equations (\ref{Jdp-1000},\ref{dpdp-1000}) imply
(\ref{dJ-1100},\ref{dpdC-1100},\ref{Jdp-1100}). 

Conversely, having (\ref{dJ-1100},\ref{dpdC-1100},\ref{Jdp-1100})
consider first the case $dC\neq 0$. Then we have from
\eqref{dpdC-1100} that $dpdp\neq 0$ and from \eqref{Jdp-1100} the
equation \eqref{Jdp-1000} follows.\footnote{Indeed, assuming that
  \eqref{Jdp-1000} is invalid we obtain from \eqref{Jdp-1100} that
  $dp$ has a rank two at most and $dpdp =0 $ contradicting
  \eqref{dpdC-1100}.} We conclude that both the system
(\ref{Jdp-1000},\ref{dpdp-1000}) and
(\ref{dJ-1100},\ref{dpdC-1100},\ref{Jdp-1100}) produce the same
solutions in the presence of the background $dC\neq 0$ or in the limit
$dC\to 0$. In the main text we considered
$dC = \a d\tilde A d\tilde A +\b \Tr (R^2)$.

% -A.4-%%%%%%%%%%%%%%%%%%%%%%%%%%%%
\subsection{Adding degrees of freedom}

One possible way to resolve the coordinate singularity for $dp=0$
is to add degrees of freedom to the theory. Let us introduce the
following gauge invariant term in the action \eqref{S-bb-4d}
\be
    \D S=\int_{{\cal M}_4} \phi dp d\tilde \t\,,
 \label{S-add}
\ee
where the new variables are the gauge invariant scalar $\phi$
and pseudo-1-form $\tilde \t$. No dynamics is introduced for them.
The equation of motion for $p_\mu$ \eqref{pi-1000} is modified as
\be
J^\mu=    [2d\psi dp +d\phi d\tilde\t]^\mu \,,
\label{J-new}
\ee
and two additional equations follow by variation with respect to the
new fields
\begin{align}
  &  d\tilde \t dp =0\,,
    \label{phi-var}
    \\
  & d\phi dp=0\,.
    \label{tau-var}
\end{align}
The equations \eqref{phi-var}, \eqref{tau-var} are trivially satisfied
in the case of interest $dp=0$, leaving completely free the new
variables $(\phi,\tilde \t)$. Then the second term in the rhs of
expression \eqref{J-new} can provide a general parameterization of the
conserved current $J$, when the first term vanishes.

The addition \eqref{S-add} remains valid in presence of gauge backgrounds,
with the replacement $p\to\pi$, and it is found that the corrections
introduced in the currents and equations of motion do not have effect
on anomalies and stress tensor conservation.  Of course the physical
consequences of including the term \eqref{S-add} into the action remain
to be fully understood.

%-A.5-%%%%%%%%%%%%%%%%%%%%%%
\subsection{Equations obeyed by \texorpdfstring{$\psi$}{Lg}}
 \label{sec:finding-psi}

Specializing \eqref{pi-1000} for nonrelativistic fluids, we obtain
\begin{align}
    &\rho = 2[d\psi d p]^0\,,
 \\
    &\rho v^i = 2[d\psi d p]^i\,,
 \\
    &[d p d p] =0\,,
\end{align}
where $ p_i =v_i$ and $ p_0 = -\mu -v^2/2$. We rewrite these
equations as
\begin{align}
    &\rho = 2\bm\nabla\psi\cdot\bm\omega\,,
    \label{rho-par}
 \\
    &\rho \bm v = -2\dot\psi\bm\omega
      +2\bm\nabla\psi\times(\dot{\bm v}-\bm\nabla  p_0)\,,
 \label{seceq}
 \\
    &\p_t(\bm v\cdot\bm\omega) +\bm\nabla\cdot(- p_0\bm\omega
      +\bm v\times(\dot{\bm v}-\bm\nabla  p_0) =0\,.
\end{align}
The last equation can be transformed into
\begin{align}
    &\bm\omega\cdot(\dot{\bm v}-\bm\nabla p_0) =0\,.
\end{align}
Vector-multiplying the second equation (\ref{seceq}) by $\bm\omega$
and assuming that $\rho>0$, we obtain
\begin{align}
    \dot{\bm v}-\bm\nabla  p_0=\bm v\times\bm\omega \,.
\end{align}
If we multiply (\ref{seceq}) by $\bm\nabla\psi$ we obtain 
\begin{align}
    \dot\psi +(\bm v\cdot\bm\nabla)\psi =0\,.
\end{align}
Collecting all equations together we write the complete system
\begin{align}
    &\rho = 2\bm\nabla\psi\cdot\bm\omega\,,
 \label{rhoinit}\\
    &\dot{\bm v}-\bm\nabla  p_0=\bm v\times\bm\omega \,,
 \label{seceq2}
 \\
    &\dot\psi +(\bm v\cdot\bm\nabla)\psi =0\,.
 \label{psitrans}
\end{align}
Let us now assume that we are given the configuration $\rho\geq 0$ and
generic $\bm v$ at $t=0$ so that $\bm\omega\neq 0$. Then, we can
locally solve (\ref{rhoinit}) and find $\psi$. After that the
equations (\ref{seceq2},\ref{psitrans}) and continuity equation for
$\rho$ (that follows from these equations) gives us the subsequent
configurations at any future time (locally).

Taking time derivative of (\ref{rhoinit}) and using
(\ref{seceq2},\ref{psitrans}) we can also derive continuity equation
\begin{align}
    \dot\rho +\bm\nabla(\rho\bm v) =0\,.
 \label{contcons}
\end{align}

This argument confirms in detail that the free variational principle
with $\psi$ is applicable for generic configuration of $(\rho,\bm v)$ and
is equivalent to that of restricted variations, provided that vorticity
$\bm \w\neq 0$.

%-B-%%%%%%%%%%%%%%%%%%%%%%
\section{Anomaly coefficients for Dirac and
  Weyl fermions }
\label{app:coeffs}

The  4+1d Abelian Chern-Simons action  is given by
\begin{align}
 S=   -\frac{1}{24\pi^2}\int_{{\cal M}_5} A  dA  dA\,.
 \label{CS100}
\end{align}
To obtain a conventional normalization we should replace
$A\to \frac{e}{\hbar c}A$. In the main text we also put
the flux quantum to one, i.e. set $1/(4\pi^2)\to 1$.
The action (\ref{CS100}) accounts for the anomaly inflow of
a single right Weyl fermion. For a set of Weyl fermions we have
\begin{align}
  S=  -\sum_i \frac{e_i^3 \chi_i}{24\pi^2}\int A_i dA_i dA_i\,,
 \label{weyls100}
\end{align}
where $e_i$, $\chi_i$ are, respectively,
charges (integer) and chiralities ($\pm 1$)
of the Weyls and $A_i$ are the gauge fields coupling to them.

From (\ref{weyls100}) we obtain the conservation of corresponding 3+1d
covariant currents $J^\mu_{i,cov}$ by differentiation and inflow
correspondence.
The consistent currents are also obtained from the associated
Wess-Zumino-Witten action. They read
\begin{align}
    \p_\mu J^\mu_{i, cov} &= - e_i^3 \chi_i \frac{1}{8\pi^2} [dA_idA_i]\,,
 \\ 
    \p_\mu J^\mu_{i, cons} &= - e_i^3 \chi_i\frac{1}{24\pi^2} [dA_idA_i]\,,
\end{align}
(we used the notation $[abcd]=\eps^{\a\b\mu\nu}a_\a b_\b c_\mu d_\nu$).

Let us now consider a set of Dirac fermions. Using the relation between
chiral and vector/axial components \eqref{chi-split}, we can rewrite
\eqref{weyls100} as
\begin{align}
  S= -\sum_i\frac{1}{4\pi^2}\int_{{\cal M}_5}
  \tilde e_i e_i^2\; \tilde A\  dA\  dA
  +\frac{\tilde e_i^3}{3}\;\tilde A  d\tilde A  d\tilde A\,.
\end{align}
Here $e_i$ and $\tilde e_i$ are (integer) vector and axial charges of
Dirac fermions, respectively. We assumed here and below that there are
two common gauge backgrounds, vector $A$ and axial $\tilde A$ acting
on all Dirac flavors. As a consequence
\begin{align}
    \p_\mu  J^\mu_{cons} &= 0\,,
 \\
  \p_\mu  J^\mu_{cov} &=
-\left(\sum_i \tilde e_i e_i^2 \right) \frac{1}{2\pi^2} [dAd\tilde A] \,,    
 \\ 
  \p_\mu \tilde J^\mu_{cons} &=
 -\left(\sum_i \tilde e_i e_i^2 \right)\frac{1}{4\pi^2} [dAdA]
-\left(\sum_i \tilde e_i^3 \right)\frac{1}{12\pi^2} [d\tilde Ad\tilde A]\,,
 \\
  \p_\mu \tilde J^\mu_{cov} &=
-\left(\sum_i \tilde e_i e_i^2 \right)\frac{1}{4\pi^2} [dAdA]
-\left(\sum_i \tilde e_i^3 \right)\frac{1}{4\pi^2} [d\tilde Ad\tilde A]\,.
\end{align}

Let us now describe the mixed axial-gravitational  anomaly,
starting with a single right Weyl fermion.
The mixed Chern-Simons gravity action is 
\begin{align}
  \D S=  -\frac{1}{192\pi^2}\int_{{\cal M}_5} \tilde A \Tr(  R^2)\,,
\end{align}
where $R=d\omega +\omega^2$ is the curvature two-form.
We obtain the following contribution to the chiral currents 
\begin{align}
    \p_\mu J^\mu_{cons} = -\frac{1}{192\pi^2}\Big[\Tr( R^2)\Big]\,,
 \\
    \p_\mu  J^\mu_{cov} = -\frac{1}{192\pi^2}\Big[\Tr( R^2)\Big]\,.
\end{align}

For a set of Weyls we have
\begin{align}
    \p_\mu J^\mu_{i, cov} &= - e_i^3 \chi_i \frac{1}{8\pi^2} [dA_idA_i]
- e_i \chi_i\frac{1}{192\pi^2}\Big[\Tr( R^2)\Big]\,, \\ 
    \p_\mu J^\mu_{i, cons} &= - e_i^3 \chi_i\frac{1}{24\pi^2} [dA_idA_i]
    - e_i \chi_i\frac{1}{192\pi^2}\Big[\Tr( R^2)\Big]\,.
\end{align}
It is easy to get from here the corresponding Dirac anomalies.

%-C-%%%%%%%%%%%%%%%%%%%%%%%%%%%%%%%%%%%%%%%
\section{Hydrodynamics in chiral backgrounds}
\label{app:C}

\subsection{Chiral topological theory and
 decoupling of 1+1 dimensional chiral currents}

We now consider a chiral background and show that 1+1d excitations
split into chiral components, one interacting with the background and
the other remaining inert.

The chiral decomposition of backgrounds and currents is defined by
\begin{align}
    &J=J_++J_-\,, \qquad \tilde J=J_+ - J_-\,,
    \qquad A=\frac{1}{2}(A_+ + A_-)\,, \qquad \tilde A=\frac{1}{2}(A_+ - A_-)\,,
 \nonumber \\
    &p=\frac{1}{2}(b_+ + b_-)\,, \qquad \tilde q=\frac{1}{2}(b_+ - b_-)\,,
 \label{split}
\end{align}
obeying the relation $J^\mu A_\mu + \tilde J^\mu \tilde A_\mu =
J^\mu_+ A_{+\mu} +  J^\mu_-  A_{-\mu}$.

We consider one chiral part by setting $A_-=b_-=0$ and redefining
$A_+=A, b_+=b$. The 2+1d topological theory \eqref{CS} becomes
\be
    S=\int \frac{1}{2} b d b + b d A\,,
\ee
leading to the Chern-Simons response action \eqref{S-CS}.

Following the steps of Section \ref{sec:2.2}, we use the anomaly inflow
to obtain the 1+1d covariant current
\be
    j^3=\frac{\d S}{\d A_3}=
    \eps^{3\mu\nu}\de_\mu b_\nu =
    \de_\a J^\a_{cov}\underset{eq.m.}{=}
    \eps^{\mu\nu}\de_\mu\left(\de_\nu \chi -A_\nu\right)\,.
 \label{chi-cur}
\ee
In the last expression, we substituted the solution of the
bulk equations of motion, leaving the gauge parameter $\chi$ free.
The WZW action obtained from the response action \eqref{S-CS} is,
\be
    S_{WZW}[\l,A]=-\frac{1}{2}\int_{{\cal M}_2} \l dA\,,
\ee
giving the anomaly of the consistent current
$\de_\a J^\a_{cons}=-\eps^{\mu\nu}\de_\mu A_\nu/2=\de_a J^\a_{cov}/2$.

The chiral splitting of excitations in the boundary Dirac theory will
be obtained on the solutions of the equations of motion. A first
indication of this decomposition comes from the form of inflow
currents \eqref{jtcov-2d}, \eqref{jcov-2d}, which we can rewrite using
\eqref{split},
\begin{align}
 \label{curr-split}
    J_{+cov}^\a&\underset{eq.m.}{=}
           \eps^{\a\b}\left(\de_\b\psi +\de_\b\th - A_{+\b}\right)\,,
\\
    J_{-cov}^\a&\underset{eq.m.}{=}
       \eps^{\a\b}\left(\de_\b\psi -\de_\b\th - A_{-\b}\right)\,.
\end{align}
One sees that for $A_-=0$, $J_{-cov}$ does not couple to the background
and vanishes for $\th=\psi$. The expression for $J_+$ matches \eqref{chi-cur}
by identifying $\chi=\psi+\th$.

It is convenient to consider the 1+1d theory in the form
\eqref{rel-hydro} used to establish duality. Paying attention to the
factor of two in the chiral splitting \eqref{split}, it can be written
\be
    S=\frac{1}{2}\int_{{\cal M}_2} - \frac{1}{2}(\pi_\mu -A_\mu)^2 +
    (A-d\psi)\pi\,.
 \label{2-act}
\ee
The equations of motion are
\begin{align}
    \pi_\mu&=\de_\mu\th\,,
 \nonumber \\
    \pi^\mu -A^\mu&=\eps^{\mu\nu}(\de_\nu\psi -A_\nu)\,.
 \label{chi-eqm}
\end{align}
The consistent current obtained from this action, evaluated on the
equations of motion, is
\be
    J^\mu_{cons} \underset{eq.m.}{=}\frac{1}{2}\epsilon^{\mu\nu}
    \left(\de_\nu\th +\de_\nu\psi -A_\nu \right)\,,
\ee
whose anomaly agree with that obtained from the above WZW action.

It remains to show that dynamics allows for a single nonvanishing
chiral current. Actually, the equations of motion \eqref{chi-eqm} in
the gauge $A_0=A_1$, became the duality relations \eqref{2d-duality}:
explicitly, $\de_0\th=\de_1\psi$, $\de_1\th=\de_0\psi$.
These admit two solutions
\begin{align}
    &\th=\psi \ \ \quad \to\quad (\de_0-\de_1)\th=0\,,
 \\
    &\th=-\psi \quad \to\quad (\de_0+\de_1)\th=0 \,.
\end{align}
The first solution gives the expected result $J_{-cov}=0$ and also
makes the other current chiral, $J_{cov}^0=-J^1_{cov}$.
The solution with opposite chirality $(\th=-\psi)$ is nonetheless present,
corresponding to $J_+=0$ and $J_-\neq 0$ but decoupled from the background.

The chiral splitting can be made more explicit by rewriting the action
\eqref{2-act} in a equivalent form by using its equations of motion
\eqref{chi-eqm}, leading to the expression,
\begin{align}
  S = \frac{1}{2}\int -\frac{1}{4}
  \p_x\theta_+(\p_t-\p_x)\theta_+
  +\frac{1}{4}\p_x\theta_-(\p_t+\p_x)\theta_-\,,
\label{chi-act}
\end{align}
where $\th_+=\th+\psi$ and $\th_-=\th-\psi$ and we set $A=0$ for simplicity.
The action has decomposed into two parts, each one describing one
chirality only, as it clear from the respective equations of motion.
Each piece is known as the chiral boson action,
because it bosonizes a single Weyl fermion \cite{arouca2022quantum}.

In conclusion, we have checked that chiral splitting occurs in 1+1 dimensions
for the free massless Dirac fermions, owing to kinematics.
In the 3+1 dimensional theory, free massless Dirac fermions also
decouple in a pair of Weyls.
However, the 3+1d hydrodynamics described in this work
does not show this feature, as explained in the next Section.

In the interacting case, there is no reason to expect the chiral
splitting of Dirac fermions in both 1+1 and 3+1 dimensions.  In the
1+1d case, one can clearly see from previous formulas that
non-quadratic Hamiltonians do not split in two independent terms.  The
fluid made out of a single copy of interacting Weyl fermions can be
described by deforming the chiral boson action \eqref{chi-act} by
adding more general Hamiltonians.

%-C.2.1-%%%%%%%%%%%%%%%%%%%%%%%%%%%%%%%%%%%%%%%

\subsection{3+1 dimensional hydrodynamics in chiral backgrounds}

The 4+1d Chern-Simons action describing the response of the system in
a chiral background and the associated 3+1d WZW action have the form
\be
  S_{ind}= -\frac{1}{6}\int_{{\cal M}_5} AdAdA \,,\qquad  \qquad
S_{WZW}= -\frac{1}{6}\int_{{\cal M}_4} \psi dAdA\,.
\ee
From the first expression and anomaly inflow follows the 3+1d 
anomaly of covariant currents, while the WZW action determines
the anomaly of consistent currents,
\begin{align}
  & * d J_{cov}= -\frac{1}{2}dAdA\,,
    \label{chi-cov-anom}
  \\
  & * d J_{cons}= -\frac{1}{6}dAdA\,, \qquad\qquad
    * J_{cov}=*J_{cons}-\frac{1}{3}AdA\,.
    \label{chi-cons-anom}
\end{align}
which have the correct values for a Weyl fermion.

The 3+1d hydrodynamics action \eqref{S-hy-psi}
is now specialized for chiral backgrounds
by setting $A=\tilde A$, i.e. $A_-=0$. For $\a=1/3$, it reads
\be
    S=\frac{1}{8}\int_{{\cal M}_4} P(p) +A\pi d(\pi+A)+
    \psi\left(d\pi d\pi+\frac{1}{3} dAdA\right)\,.
 \label{chi-act-5d}
\ee
The overall normalization has been changed as explained in the following. 
Upon variation, one obtains the equations of motion 
\be
-\frac{\de P}{\de \pi_\nu}=\left[
 (\pi-A)dA +2(d\psi-A)d\pi\right]^\nu,
\qquad
d\pi d\pi+\frac{1}{3}dAdA=0\,,
\label{chi-eq-m}
\ee
and the consistent current
\begin{align}
J_{cons}^\nu& =-\frac{1}{8}\frac{\de P}{\de \pi_\nu} +\frac{1}{8}\left[
 \pi d(\pi+A)+d(A\pi) +\frac{2}{3}d\psi dA\right]^\nu
  \\
 &\underset{eq.m.}{=}\frac{1}{8}\left[
   (\pi-A) d(\pi+A)+2 d(A\pi) +2 d\psi d(\pi+\frac{1}{3} A)\right]^\nu,
   \label{chi-curr-cons}
  \end{align}
One verifies that this current obeys the anomaly
\eqref{chi-cons-anom}, thus checking the normalization.

The covariant current can also be found from $J_{cov}+\tilde J_{cov}$
of the non-chiral theory \eqref{jcov4d-eqm}, \eqref{jtcov4d-eqm}
evaluated for $A=\tilde A$, up to normalization,
\be
*J_{cov}\underset{eq.m.}{=}
\frac{1}{8}\left[ (\pi-A)d(\pi+A) +2(\a+1)(d\psi-A)dA
  -2d((d\psi-A)(\pi-A))\right]\,.
\label{chi-curr-4d}
\ee
For $\a=1/3$, this expression is indeed equal to \eqref{chi-curr-cons}
plus correction \eqref{chi-cons-anom}, as expected; it
shows gauge invariance, owing to simultaneous invariance of the
$(d\psi-A)$ and $(\pi-A)$ terms.

%-6-%%%%%%%%%%%%%%%%%%%%%%%%%%%%%%%%%%%%%%%

\subsubsection{Absence of chiral decoupling in 3+1 dimensions}

We now ask whether the chiral currents
$J_{\pm cov}=J_{cov} \pm \tilde J_{cov}$ decouple
in the chiral background $A=\tilde A$, i.e. $A_-=0$.
In this case, chirality is not a simple kinematic condition.

If e.g. the right $(+)$ Weyl is probed by a gauge background and the left
$(-)$ one is not,
we would expect that there are solutions of the equations of motion with
vanishing $(-)$ current.
The current $J_{+cov}=J_{cov}+ \tilde J_{cov}$ was found before in
\eqref{chi-curr-4d}. Next we check whether the difference of currents
$J_{-cov}=J_{cov}- \tilde J_{cov}$
can vanish, thus proving chiral splitting in the 3+1d Dirac hydrodynamics.

From \eqref{jcov4d-eqm}, \eqref{jtcov4d-eqm}, we find (for
$\a=1/3$):
\begin{eqnarray}
\!\!\!\!\!\!\!  *\tilde J_{-cov} &\underset{eq.m}{=}&
\pi d\pi +\frac{1}{3} AdA -d(A\pi) +\frac{2}{3}d\psi dA -2 d\psi d\pi
\nonumber          \\
&=& (\pi -d\psi)d\pi +\frac{1}{3}(A-d\psi)dA-
    d[(d\psi -A)(d\psi -\pi)]   \label{diff-curr}\\
&=& dX \neq 0\,.
\nonumber
\end{eqnarray}

In these relations, we used the first equation of motion \eqref{chi-eq-m}
but not the second one $d\pi d\pi+dAdA/3=0$.

We remark that in chiral backgrounds the current \eqref{diff-curr}
has no anomaly, as verified by using the remaining equation of motion.
It follows that  \eqref{diff-curr} actually is an exact form
$dX$. However, it does not vanish as a consequence of the equations
of motion, not even for $A=0$. Therefore, the two chiralities are
always coupled.

Let us remark that a physical instance of a single Weyl  
is at the boundary of the 4+1d topological insulators, so it should be
possible to obtain its hydrodynamic description by the methods
outlined in this work. Presumably one should start from the `two-fluid'
4+1d topological theory described in Section \ref{sec:431} and decouple it
in two symmetric parts.

\bibliographystyle{JHEP}
\bibliography{hydro-anomalies}

\end{document}